\documentclass[11pt, leqno]{article}
\usepackage[english]{babel}
\usepackage{latexsym}
\usepackage{amssymb} 
\usepackage{amsmath} 
\usepackage{graphics}
\usepackage{amsmath} 
\renewcommand{\baselinestretch}{1.5} 
\numberwithin{equation}{section}
\usepackage{harvard}

\newtheorem{theorem}{Theorem}[section]
\newtheorem{lemma}[theorem]{Lemma}
\newtheorem{corollary}[theorem]{Corollary}
\newtheorem{proposition}[theorem]{Proposition}
\newtheorem{definition}[theorem]{Definition}
\newtheorem{example}[theorem]{Example}
\newtheorem{remark}[theorem]{Remark}
\newcommand{\ve}{\varepsilon}

\newcommand{\mc}{\mathcal}

\begin{document}
\renewcommand{\baselinestretch}{1.5} 
\title{Cash Sub-additive Risk Measures and Interest Rate Ambiguity\thanks{
Address correspondence to Claudia Ravanelli, Swiss Banking
Institute, University of Zurich, Plattenstrasse~14, CH-8031 Zurich,
Switzerland. E-mail: \texttt{ravanelli@isb.uzh.ch}.
E-mail address for Nicole El Karoui:
\texttt{elkaroui@cmapx.polytechnique.fr}.
For valuable comments we thank
seminar participants at SAMSI 2006 workshop, Evry Workshop 2006,
Princeton 2006 conference on Risk Measures and Robust Control in
Finance and Cass Business School.
We are grateful to the anonymous referees for many useful and
fruitful suggestions which improved the paper.
Ravanelli's research was supported by the University Research
Priority Program ``Finance and Financial Markets'' (University of
Zurich) and by the NCCR-FinRisk (Swiss National Science Foundation).
} }
\author{Nicole El Karoui$^{\mbox{\scriptsize a}}$ \hspace{2cm}
Claudia Ravanelli$^{\mbox{\scriptsize b}}$\\
$^{\mbox{\scriptsize a}}${CMAP, Ecole Polytechnique, France}\\
$^{\mbox{\scriptsize b}}${Swiss Banking Institute, University of
Zurich, Switzerland}}
\maketitle
\!\!\!\!\!
\begin{abstract}
A new class of risk measures called cash sub-additive risk
measures is introduced to assess the risk of future financial,
nonfinancial and insurance positions.
The debated cash additive axiom is relaxed into the cash
sub-additive axiom to preserve the original difference between the
num\'{e}raire of the current reserve amounts and future positions.
Consequently, cash sub-additive risk measures can  model stochastic
and/or ambiguous interest rates or defaultable contingent claims.
Practical examples are presented and in such contexts cash
additive risk measures cannot be used.
Several representations of the cash sub-additive risk measures are
provided. The new risk measures are characterized by penalty
functions defined on a set of sub-linear probability measures and
can be represented using penalty functions associated with cash
additive risk measures defined on some extended spaces.
The issue of the optimal risk transfer is studied in the new
framework using inf-convolution techniques.
Examples of dynamic cash sub-additive risk measures are provided via
BSDEs where the generator can locally depend on the level of the
cash sub-additive risk measure.
\end{abstract}

\newpage
\renewcommand{\baselinestretch}{2} 
\section{Introduction}
The assessment of financial and nonfinancial risks plays a key role
for economic agents when pricing assets or managing their wealths.
Consequently, over the last decade several measures of risk have
been proposed to assess the riskiness of financial and nonfinancial
positions and compute cash reserve amounts for hedging purposes.
The axiomatic based monetary risk measures have been largely
investigated because most axioms embed desirable economic
properties.
Coherent risk measures have been introduced by
\citeasnoun{artzner_et_al_1997}, \citeasnoun{artzner_et_al_1999},
and further developed by \citeasnoun{delbaen_2000},
\citeasnoun{delbaen_2002};
sublinear risk measures by \citeasnoun{frittelli_2000b};
convex risk measures by \citeasnoun{foellmer_schied_2002},
\citeasnoun{foellmer_schied_2002b} and
\citeasnoun{frittelli_gianin_2002}.
Examples of convex risk measures related to pricing and hedging in
incomplete markets are  provided by, for instance,
\citeasnoun{elkaroui_quenez_1996},
\citeasnoun{carr_geman_madan_2001},
\citeasnoun{frittelli_gianin_2004} and \citeasnoun{staum_2004}.
However, while the convexity and the monotonicity axioms have been
largely accepted by academics and practitioners, the cash additive
axiom has been criticized from an economic viewpoint.
A basic reason is that while regulators and financial institutions
determine and collet today the reserve amounts to cover future
risky positions, the cash additivity requires that risky positions
and reserve amounts are expressed in the same num\'{e}raire.
This is a stringent requirement that limits the applicability of
cash additive risk measures.
Implicitly it means that risky positions are discounted before
applying the risk measure assuming that the discounting process does
not involve any additional risk.
Unfortunately, when the interest rates are stochastic this
procedure does not disentangle the risk of the financial position
per s\'{e} and the risk associated to the discounting
process\footnote{Disentangling the different risks is crucial when
implementing hedging strategies as different risks are hedged on
different markets.}.
Furthermore, payoff functions on risky assets are a priori and
contractually determined by economic agents considering different
scenarios for the underlying asset.
While this procedure is theoretically framed into cash additive risk
measures, the cash additive axiom does not allow to account for
ambiguous discount factor.
For a correct assessment of the current reserve amount it is equally
important to allow for ambiguity on the underlying asset and on the
discount factor.
This assessment is achieved by relaxing the cash additive axiom and
searching for risk measures that preserve the different
num\'{e}raires of the current reserve amounts and the future risky
positions.

The main contribution of this paper is to propose a new class of
risk measures called cash sub-additive risk measures 
that are directly defined on the future risky positions and provide
the reserve amounts in terms of the current num\'{e}raire.
To reconcile the two different num\'{e}raires cash sub-additive risk
measures relax the cash additive axiom into the cash sub-additive
axiom. This is the minimal requirement to account for the time value
of money.
Remarkably, the cash sub-additive axiom is enough  to characterize
measures of risk that can be applied also when the cash additive
risk measures cannot---as for instance under ambiguous interest
rates or defaultable cash flows.
Cash sub-additive risk measures turn out to be suitable not only for
assessing financial risks but also insurance and other kind of risks.
For example, the put option premium investigated by
\citeasnoun{jarrow_2002} as a measure of the firm insolvency risk
defines a cash sub-additive risk measure.
Moreover, similarly to the cash additive risk measures, the cash
sub-additive risk measures can be represented using penalty
functions. In particular, we show that cash sub-additive risk
measures are characterized by minimal penalty functions which only
depend on finitely additive set functions $\mu$ such that $0 \leq
\mu(\Omega) \leq 1$, that we call finitely additive sub-probability
measures.


The other contributions of this paper are the following.
In the framework of cash additive risk measures when the zero-coupon
bond is available for the relevant time horizon, we provide the
conditions under which discounting the forward risk measure to
obtain current reserve amounts defines risk measures additive with
respect the current num\'{e}raire and vice versa
(Section~\ref{sec:spot_forward_risk_measures}).

In Section~\ref{sec:cash_sub_additive_risk_measures} we introduce
the cash sub-additive risk measures that we denote by $\mc R$.
We provide several examples of these new risk measures that
generalize the put option premium and naturally arise when
accounting for ambiguous discount factor or insurance risks.
These risk measures are obtained composing cash additive risk
measures and a specific class of random convex functions.
A representation result showing the impact of the ambiguous discount
factor/num\'{e}raire is given.

In Section~\ref{sec:cash_sub_additive_dual_reperesentation} we study
the dual representation of cash sub-additive risk measures. Instead
of using convex analysis tools, we extend cash sub-additive risk
measures to an enlarged space of risky positions where they become
cash additive.
This approach provides a rich financial
interpretation of both cash additive and cash sub-additive risk
measures and allows to derive properties of $\mc R$ using the classical
theory on cash additive risk measures.
Using the duality result, a characterization of $\mc R$ in terms of
deterministic discount factors is easily obtained where any cash
sub-additive risk measure can be represented as the worst case
scenario of a family of discounted forward risk measures.

In Section~\ref{sec:cash_sub_additive_credit_risk} two other links
between cash sub-additive and cash additive risk measures are
presented where more involved techniques are required.
The first link indicates a possible way to recover a representation
of a general cash sub-additive risk measure where the ambiguous
num\'{e}raire is explicitly modeled as a random variable on the
original space of definition of $\mc R$.
The second link shows that cash sub-additive risk measures generated
via convex  functions are compositions of an unconditional and a
conditional cash additive risk measures.

In Section~\ref{sec:Optimal derivative design and inf-convolution}
using  cash sub-additive risk measures we study the problem of
designing the optimal transaction between two economic agents  in a
general framework allowing for ambiguous discount factors. In
particular we show that the risk transfer problem can be reduced to
an inf-convolution of cash sub-additive risk measures which is again
a cash sub-additive risk measure.

Finally, in Section~\ref{sec:infinitesimal of cash sub-additive risk
measures} we provide a dynamic example of cash sub-additive risk
measures which are solutions of backward stochastic differential
equations (BSDEs).
In contrast to the cash additive risk measures generated via BSDEs,
the generator of dynamic cash sub-additive risk measures,
besides being a function of the martingale part, can also depend on
the level of the cash sub-additive risk measure, generating recursive risk
measures.
Section~\ref{sec:conclusion} concludes.
%
%
%
%

\section{Cash additive risk measures}\label{sec:cash_invariant_risk_measures}
In this section we recall some key properties of cash additive risk
measures and we discuss the cash additive axiom. The following
definitions are consistent with the definitions of monetary risk
measure in \citeasnoun{foellmer_schied_2002b}.
\subsection{Definitions and properties of cash additive risk measures}
Let $(\Omega, \mc{A})$ be a measurable space. The risky positions at
the relevant time horizon belong to the linear space of bounded
functions including constant functions denoted by $\mc{X}$.
\begin{definition}\rm
{\it A cash additive risk measure} is a functional $\rho
:\mc{X}\rightarrow \mathbb{R}$ cash additive,
convex and monotone decreasing, i.e.,\\
$a)$ Convexity: $\forall \lambda \in [ 0,1], \quad \rho\big(
\lambda X+( 1-\lambda ) Y\big) \leq \lambda \rho(X) +( 1-\lambda
) \rho(Y) $;\\
$b)$ Monotonicity: $X \leq Y \Rightarrow \rho(X) \geq \rho (Y)$;\\
$c)$ Cash additivity (or cash invariance): $\forall m\in \mathbb{R},
\quad \rho(
X+m) =\rho(X) -m$. \\
A  cash additive risk measure is {\it coherent} when \\
$d)$ Positive homogeneity: $\forall \lambda \in \mathbb{R}^+,
\quad \rho\big(
\lambda\,X\big)=\lambda\rho\big(X\big)$.\\
$e)$ $\rho$ is {\rm normalized} when $\rho(0)=0$.
\\
$f)$  $\rho$ is {\rm continuous from below (from above)} when
\begin{eqnarray*}
X_n \nearrow X  \,\,\,\, \Rightarrow \,\,\,\, \rho(X_n)\searrow
\rho(X),\  \  \  \  (X_n \searrow X  \,\,\,\, \Rightarrow \,\,\,\,
\rho(X_n) \nearrow \rho(X)).
\end{eqnarray*}
\end{definition}
The convexity axiom translates the natural important fact that
diversification should not increase risk. In particular, convex
combinations of ``admissible'' risks should be ``admissible''.
%
\\
%
To shorten the representation of convex combinations of elements we
use the following notation. We denote the {\it{barycenter}} (or
{\it{convex combination}}) of the set  $x_I:=\{x_{(1)}, x_{(2)},
\ldots, x_{(I)}\},~I\in \mathbb{N}$,
\begin{eqnarray}\label{baricenter}
{Bar}[x_I]:={Bar}^{\lambda_I}[x_I]:= \sum_{i=1}^I \lambda_i x_{(i)}
\mbox{ where } \lambda_i \in [0,1],\, i=1,\ldots,I,  \mbox{ and }
\sum_{i=1}^I \lambda_i = 1.
\end{eqnarray}
In particular, $f$ is a convex function if and only if
$f({Bar}[x_I])\leq {Bar}[f(x)_I]$. The same definition holds for a
set $X_I$ of random variables.

\subsection{Dual representation of cash additive risk measures}
A key property of cash additive risk measures is the dual
representation in terms of normalized finitely additive set
functions and minimal penalty functional (\citeasnoun[Theorem
4.12]{foellmer_schied_2002b}).
The dual point of view emphasizes the interpretation in terms of a
worst case scenario related to the agent's (or regulator's) beliefs:
the agent does not know the true ``probability'' measure and uses
distorted beliefs from a subjective set of normalized additive  set
functions.
%
Under the additional assumption that risk measures are continuous
from below, the dual representation is in  term of $\sigma$-additive
probability measures (\citeasnoun[Proposition
4.17]{foellmer_schied_2002b}).
\begin{theorem}\label{Theorem dual representation}
(a) Let $\mc{M}_{1,f}(\mc{A})$ be the set of all finitely additive
set functions $Q$ on $(\Omega,\mc{A})$ normalized to one,
$Q(\Omega)=1$, and $\alpha$ the minimal penalty functional taking
values in $\mathbb{R}\cup \big\{ +\infty \big\}$:
\begin{eqnarray}\label{alpha_min}
& & \forall Q\in \mc{M}_{1,f}(\mc{A}), \quad \alpha(Q) = \sup_{X \in
\mc X}\big\{ \mathbb{E}_{Q} [-X] -\rho ( X) \big\}, \qquad \big(
\geq -\rho( 0) \big)
\\
& & \mc Dom(\alpha)= \{Q \in \mc {M}_{1,f}(\mc{A})| \ \alpha \big(
Q \big)<+\infty\}.
\end{eqnarray}
The Fenchel duality relation holds:
\begin{eqnarray}
& & \forall X \in \mc{X, \quad }\rho( X) =\sup_{Q%
\in \mc{M}_{1,f}(\mc{A})}\big\{ \mathbb{E}_{Q}[-X] -\alpha \big(Q)
\big\}.
\end{eqnarray}
Moreover, for any $X \in \mc X$ there exists a $Q_X \in \mc
M_{1,f}(\mc{A})$, such that
$\rho( X) =\mathbb{E}_{Q_X}[-X] -\alpha \big( Q_X\big)= \max_{Q \in
\mc{M}_{1,f}(A)}\big\{ \mathbb{E}_{Q}[-X] -\alpha(Q) \big\}. $\\
%
(b) Let $\mc M_1(\mc A)$  denote the set of all probability measures
$\mathbb{Q}$  on $(\Omega, \mc A)$.
Let $\rho$ be a monetary risk measure continuous from below (Fatou
property) and suppose that $\beta$ is any penalty function on $\mc
M_{1,f}(\mc A)$ representing $\rho$. Then $\beta$ is concentrated on
the class $\mc M_1(\mc  A)$ of probability measures, i.e.,
$\beta(Q)<\infty$ only if $Q$ is $ \sigma\mbox{-additive}.$
\end{theorem}
See \citeasnoun{kratschmer_2005} for necessary
conditions to obtain representation results in terms of probability
measures.

The following lemma shows that a cash additive risk measure is
linear with respect to the linear subspace generated by a position
$Y$ if and only if any $Q$ in the domain of the penalty functional
satisfies the calibration constraint: $Q(-Y)=\rho(Y)$. This lemma
will be used to derive the results in
Section~\ref{sec:spot_forward_risk_measures}.

\begin{lemma}\label{homogeneity_linearity}
Let $\rho$ be a normalized cash additive risk measure on $\mc X$ and
$\mc W$ a linear subspace of $ \mc X$ containing the constants. The
risk measure $\rho$ is a linear on $\mc W$,
if and only if $\rho(W )= \mathbb{E}_{Q}[-W]$ for any $Q \in \mc
Dom(\alpha)$. This implies that the risk measure is invariant with
respect to $\mc W$, that is $\forall X \in \mc X,\,\forall W \in \mc
W,\, \rho(X+ W)=\rho(X) + \rho( W).$
\end{lemma}
\textit{Proof.}  The dual representation  and the linearity of
$\rho$ with respect to $\mc W$ imply that for any $Q \in \mc
Dom(\alpha)$, $\lambda \in \mathbb{R}$,
$\lambda \rho(W)= \rho(\lambda W) \geq \mathbb{E}_{Q} [\lambda (-
W)]- \alpha(Q)$,  where $\alpha $ is  the minimal penalty of $\rho$.
Then $ \alpha(Q) \geq -\lambda \left( \rho(W)+
\mathbb{E}_{Q}[W]\right)$.
As the last inequality  holds for any $\lambda \in \mathbb{R}$,
$\rho(W)= -\mathbb{E}_{Q} [W],\ \ \forall Q \in \mc Dom(\alpha).$
The vice versa is evident.

If the calibration constraint holds, then $\rho(X+W)= \sup_{Q \in
\mc Dom(\alpha)} \{\mathbb{E}_Q\big[-X- W\big ]- \alpha(Q)\}=\sup_{Q
\in \mc Dom(\alpha)} \{\rho(W) +\mathbb{E}_Q[-X]-
\alpha(Q)\}=\rho(W) +\rho(X), $ for any $X \in {\mc X},\, W \in {\mc
W}.$ $\hfill{\Box}$

\subsection{Cash additivity and discounting}
The cash additive axiom is motivated by the interpretation of
$\rho(X)$ as capital requirement\footnote{See
\citeasnoun{frittelli_scandolo_2005} for an extensive study of the
axiom of cash additivity and the related concept of capital
requirement.}. Intuitively, $\rho(X)$ is the amount of cash which
has to be added to the risky position $X$ in order to make it
acceptable (i.e., with non positive measure of risk) by a
supervising agency
$$ \rho(X+\rho(X))=\rho(X)-\rho(X)=0.$$
Hence the cash additive property requires that the risky position
and the risk measure are expressed in the same num\'eraire.
Then either cash additive risk measures are defined on the
discounted value of the future positions (see, for instance,
\citeasnoun{delbaen_2000} and \citeasnoun{foellmer_schied_2002b})
or cash additive risk measures are defined directly on the future
positions and give the forward reserve amount to add to the future
position at the future date (see, for instance,
\citeasnoun{rouge_elkaroui_2000}).
In the next section, assuming that all the agents use the same
discount factor for the maturity of interest and there exists a zero
coupon bond for that maturity, we provide a link between cash
additive risk measures on the discounted positions and forward cash
additive risk measures.

In the sequel, $(\Omega, \mc{F}_T)$ is a measurable space and the
risky position belongs to $\mc{X}$, the linear space of real-valued
bounded random variables on $(\Omega, \mc{F}_T)$ including
constants.
The riskiness of $X_T \in \mc X$ is assessed at time $t=0$ and $1_T$
denotes one unit of cash available at date $T$.
$D_T$ is the stochastic (non-ambiguous) discount factor for the
maturity $T$ used by all agents in the market.
When available on the market, $B_{0, T}>0$  denotes the price at time $t=0$ of a
zero coupon bond that pays $1$ unit of cash at time $t=T$.

\subsection{Forward  and spot risk measures under stochastic discount
factor}\label{sec:spot_forward_risk_measures}
The following definitions of risk measures highlight with respect to
which num\'{e}raire the risk measures are cash additive.
\begin{definition}\rm
{\it{a)}} Let $D_T$ be the $\mc{F}_T$-measurable, $0\leq D_T\leq 1$,
discount factor  used by all agents in the market.
A {\it{spot risk measure}}, $\rho_{0}$, is a cash additive risk
measure defined on the discounted position $D_T X_T, \,\, X_T \in
\mc X$. The spot cash additive property is with respect to the cash
available at time $t=0$, $\forall X_T \in \mc X$,
\begin{equation}\label{spotrm}
\forall m\in \mathbb{R}, \  \   \rho_0 \big({D_T}X_T + \,m \big) =
\rho_0({D_T} X_T)+ \rho_0(m ) \> \mbox{ and } \>
 \rho_0(m)= m \rho_0(  1) =-m.
\end{equation}
{\it{b)}} A {\it{forward risk measure}}, $\rho_T$, is  a cash
additive risk measure defined on the future position $X_T \in
\mathcal{X}$. The forward cash additive property is with respect to
cash available at time $T$, $\forall X_T \in~\mc X$,
\begin{equation}\label{forwardrm}
\forall m\in \mathbb{R}, \  \   \rho_T(X_T + m 1_T) =
\rho_T(X_T)+\rho_T(m 1_T )\> \mbox{ and } \> \rho_T(m 1_T )=
m\rho_T( 1_T )=-m   1_{T}.
\end{equation}
\end{definition}
The spot risk measure $\rho_0$ is the monetary risk measure defined
in \citeasnoun{foellmer_schied_2002b}. It represents the cash amount
at $t=0$ to add to the discounted position $D_T X_T$ to make it
acceptable.
The spot risk measure does not disentangle the discounting risk from
the risk of the financial position per s\'{e}. Furthermore, to
meaningful consider the discounted future position the discount
factor cannot be ambiguous.
The forward risk measure $\rho_T$  gives the forward cash amount
(evaluated at $t=0$) to add at $t=T$ to the position to make it
acceptable.
When the zero coupon bond $B_{0,T}$ is available, the forward
reserve  $\rho_T(X_T)$ can be easily discounted at $t=0$.
The following proposition shows that this procedure defines a spot
risk measure when $\rho_T$  satisfies a calibration constraint on
$D_T$ and $B_{0,T}$.
Similarly, the spot risk measure $\rho_0$ capitalized by $B_{0,
T}^{-1}$ defines a forward risk measure if $\rho_0$ satisfies a
similar calibration constraint on $D_T$ and $B_{0,T}$.
The penalty function of $\rho_0$ is equal to the penalty function of
$\rho_T$ discounted by $B_{0,T}$ and the corresponding additive set
functions satisfy the usual spot-forward change of measure.

\begin{proposition}\label{CNSfromforwardtospot}
\textit{1)} Let $\rho_0$ be a normalized spot risk measure with
minimal penalty function $\alpha_0$. The functional
\begin{equation}\label{forwardrm_from_spotrm}
q_T \big( X_T \big): = B_{0,T}^{-1} \rho_0(D_TX_T), \  \  X_T\in \mc
X,
\end{equation}
is convex and monotone decreasing with respect to $X_T$, and forward
cash additive if and only if $\rho_0$ satisfies the calibration
constraint,
$ \forall \lambda \in \mathbb{R},\  \rho_0 \left(\lambda
{D_T}\right) = - \lambda B_{0,T}.$
In this case, any $Q_0 \in {\mc Dom}(\alpha_0)$ is such that
$\mathbb{E}_{Q_0}[{D_T}]=-B_{0,T}$.
Moreover, if $D_T$ is bounded away from 0, $q_T$ satisfies the
calibration constraint,
$\forall \lambda \in \mathbb{R}, \   q_T(\lambda D_T^{-1} )=
B_{0,T}^{-1} \rho_0(\lambda) =-\lambda \,{B_{0,T}}^{-1} =\lambda
q_T(D_T^{-1}),$
and the minimal penalty functional of $q_T$, $\alpha_T$, is given~by
\begin{equation}\label{ciao}
\alpha_T ({Q}_T) = B_{0,T}^{-1} \, \alpha_0(Q_0), \, \forall {Q}_T:\
dQ_0 = \frac{B_{0,T} }{ D_T}\>dQ_T \> \in {\mc Dom}(\alpha_0),\
\mbox{ and } \alpha_T=\infty \mbox{ otherwise.}
\end{equation}
\textit{2)} Let the discount factor $D_T$ be bounded away from 0.
Given  a normalized forward risk measure $\rho_T$ with penalty
function $\alpha_T$, the functional
\begin{equation}\label{spotrm_from_forwardrm}
\, q_0\big( Y \big): = B_{0,T} \rho_T(YD_T^{-1}),\  \  Y\in \mc X
\end{equation}
is convex and monotone decreasing with respect to $Y$ and satisfies,
$\forall \lambda \in
 \mathbb{R},\  q_0(\lambda D_T )= B_{0,T} \rho_T(\lambda) = -\lambda
\,{B_{0,T}} =\lambda q_0(D_T). $ Moreover, $q_0$ is a spot risk
measure if and only if $\rho_T$ satisfies: $\rho_T (\lambda
{D_T}^{-1}) = - \lambda {B_{0,T}}^{-1}=\lambda \rho_T({D_T}^{-1}),
\forall \lambda \in \mathbb{R}.$
\end{proposition}
\textit{Proof.} \textit{1)} If $\rho_0$ satisfies $\rho_0
\left(\lambda {D_T}\right) = - \lambda B_{0,T},  \forall \lambda \in
\mathbb{R},$ the forward cash additivity of $q_T$ follows directly
from Lemma \ref{homogeneity_linearity}.
Conversely, let $q_T$ be cash additive. This is equivalent to
require that $\rho_0$ satisfies
\begin{equation}\label{CNSforward-cash-invariance}
\forall X_T \in \mc X, \forall \lambda \in \mathbb{R},\  \ \ \rho_0
\left(D_TX_T + \lambda 1_T\,{D_T}\right)= \rho_0(D_TX_T)-
\lambda\,{B_{0,T}}.
\end{equation}
Setting $X_T=0$ in (\ref{CNSforward-cash-invariance}) gives the
result.
To prove (\ref{ciao}) we observe that if  $q_T$ in
(\ref{forwardrm_from_spotrm}) is a spot risk measure with minimal
penalty function $\alpha_T$, the definition of the minimal penalty
function and Lemma \ref{homogeneity_linearity} give
\begin{equation*}
\alpha_T(Q_T)=  \sup_{X_T}\{\mathbb{E}_{Q_T}[ -X_T] - q_T(X_T)\}
              =  \sup_{X_T}
              \{B_{0,T}^{-1}\mathbb{E}_{Q_0}[-D_TX_T]-B_{0,T}^{-1}\rho_0(D_TX_T)\}.
\end{equation*}
Since $D_T$ is bounded away from $0$, the one to one correspondence
between bounded variables and discounted bounded variables implies
$\alpha_T(Q_T)={B_{0,T}}^{-1}\alpha_0(Q_0)$, where $dQ_0=
{D_T}^{-1}{B_{0,T}}dQ_T$.
It follows that $Q_T$ is in the domain of $\alpha_T$ if and only if
$Q_0$ is a set function in the domain of $\alpha_0$ and satisfies
the calibration constraint in (\ref{ciao}).
Conversely, a risk measure with minimal penalty functional
$\alpha_T$ satisfying (\ref{ciao}) is  of the  form $\rho_T(
X_T)=B_{0,T}^{-1}\>\rho_0(D_TX_T)$.\\
\textit{2)} Similar arguments can be used to prove the vice versa.
$\hfill{\Box}$\vspace{2mm}

Unfortunately, the procedure of computing current reserve amounts
discounting forward risk measures (given by $q_0$ in equation
(\ref{spotrm_from_forwardrm})) is feasible only when the zero coupon
bonds for the relevant maturities are available on the market.
In this case the functional $q_0$ in equation (\ref{spotrm_from_forwardrm})
is an example of the general capital requirement defined in
\citeasnoun{frittelli_scandolo_2005}.

Next section contains the major contribution of this paper which is
the introduction of a new class of risk measures called cash
sub-additive risk measures. These risk measures provide reserve
amounts which account for the ambiguity on the discount factor. This
result is achieved by simply relaxing the cash additive axiom into
the cash sub-additive axiom and preserving the original difference
in the num\'{e}raires of reserves and future positions. This  will
be illustrated by several examples in the finance and insurance
frameworks.

\section{Cash sub-additive risk measures}\label{sec:cash_sub_additive_risk_measures}
The following observation provides the intuition for introducing
cash sub-additive  risk measures.
Given the (stochastic) discount factor $0\leq D_T\leq1$ and a spot
risk measure $\rho_0$ in equation (\ref{spotrm}), the convex,
non-increasing functional defined on $\mathcal{X}$ denoted by
$\mathcal{R}(X_T)= \rho_0(D_T\,X_T)$ is cash {\em sub-additive},
that is, it satisfies the following inequality: $\forall m \geq 0$,
$${\mc R}(X_T +m 1_T) = \rho_0(D_T\,X_T+  D_T \,m ) \geq \rho_0(
D_T\,X_T + m) = \rho_0(D_T\,X_T)-m =\mc R(X_T )-m.$$
This inequality is a simple consequence of the time value of the
money, i.e.\ $D_T \,m \leq m$.
The functional $\mc R$ is expressed in terms of the current
num\'{e}raire but directly defined on the future position expressed
in terms of the future num\'{e}raire. The function $m \in
\mathbb{R}\mapsto \mc R(X_T 1_T +m 1_T)+ m $ is non-decreasing, that
is $\mc{R}$ is cash sub-additive.
This observation highlights the cash sub-additive axiom as the
minimal condition (imposed by the time value of the money) that has
to be satisfied by risk measures which preserve the two different
num\'{e}raires of current reserve amounts and future risky
positions.
Remarkably, replacing the cash additive axiom with the cash
sub-additive axiom is sufficient to characterize risk measures  that
can be used also when cash additive risk measures cannot. For
instance under stochastic and/or ambiguous interest rates or
assessing the risk of defaultable contingent claims.
In the sequel we formally define the cash sub-additive risk
measures denoted by $\mc R$. Then we provide several examples
showing the different applications of  these new risk measures.
The previous considerations and the following examples motivate the
study of cash sub-additive risk measures.

\subsection{Definition of cash sub-additive risk measures}
\begin{definition}\label{R}\rm
A \textit{cash sub-additive risk measure} $\mc R $ is a functional
$\mc R: \mc X \rightarrow \mathbb{R}$, convex and non increasing
satisfying the cash sub-additive axiom:\\ \centerline{$
 \forall m \>\in \mathbb{R}, \,\,\mc R (X_T + m 1_T)+m \mbox{ is non
decreasing in } m. $}\\
A cash sub-additive risk measure $\mc R $ is \textit{coherent} when
$\mc R(\lambda X)=\lambda \mc R(X)$, $\forall \lambda\geq 0$.  A
cash sub-additive risk measure $\mc R $ is \textit{normalized} when
$\mc R(0)=0$.
\end{definition}
The cash sub-additive axiom can also be expressed:
\\ \centerline{$\forall
m \in \mathbb{R}, \ \ \mc R (X_T + |m| 1_T) \geq \mc R(X_T)-|m| \
\ \mbox{ and }\ \
 \mc R (X_T - |m| 1_T) \leq \mc R (X_T) +|m|.$}
Cash sub-additive risk measures naturally account for the time
value of  money. When $m$ dollars are added to the future position
$X_T$, $X_T+m 1_T$, the capital requirement at time $t=0$ is
reduced by less than $m$ dollars, that is $\mc R(X_T1_T+ m1_T)
\geq \mc R(X_T1_T) - m$.

\subsection{Examples of cash sub-additive risk measures}\label{Examples}
This section provides several examples of cash sub-additive risk
measures. All these risk measures can be obtained composing cash
additive risk measures and convex real (random) functions.
The first example arises naturally considering an ambiguous
discount factor.

\subsubsection{Cash sub-additive risk measures under ambiguous discount factors}\label{subsubsec:csarm_ambiguous_discount}
Consider a regulator assessing the risk of a future payoff $X_T$
when the discount factor $D_T$ is ambiguous and ranges between two
positive constants, $0\leq d_L \leq D_T \leq d_H \leq 1$, according
to her beliefs.
The regulator is endowed with a spot risk measure ${\rho_0}$ and
adverse to ambiguity on discount factor. Hence she assesses the
risk of $X_T$ in the interest rates worst case scenario
\begin{equation}\label{example:csarm_ambiguous_discount}
\mc R^{{\rho_0}}(X_T):= \sup_{D_T \in \mc X }\big\{
{\rho_0}(D_T\,X_T)\,|\, d_L \leq D_T \leq d_H \big\}.
\end{equation}
\begin{proposition}\label{prop:csarm_ambiguous_discount}
The functional $\mc R^{{\rho_0}}$ in equation
(\ref{example:csarm_ambiguous_discount}) is a cash sub-additive risk
measure. $\mc R^{{\rho_0}}$ can be rewritten as $\mc
R^{{\rho_0}}(X_T) ={\rho_0}(-v(X_T))$, where $v(x) =-( d_L x^+ \,-
d_H\,x^{-})$ is convex decreasing function with left derivative
$v_x$ such that $v_x \in [-1, 0]$ and $x^+=\sup(x,0)$,
$x^-=\sup(-x,0)$.
\end{proposition}
\textit{Proof.} $\mc R^{{\rho_0}}$ is a cash sub-additive risk
measure as it is the supremum of  cash sub-additive, convex and
monotone functions with respect to $X_T \in \mc X$.
Moreover, since the $\inf_{D_T \in \mc X }\{D_T\,X_T| d_L \leq D_T
\leq d_H\}$ is attained by $D^*_T=d_L {1}_{\{X_T\geq 0\}}+d_H
{1}_{\{X_T < 0\}}$, then
$\sup_{D_T \in \mc X }\big\{ {\rho_0}(D_T X_T)| d_L \leq  D_T \leq
d_H\big\} = \rho_0\big(\inf_{D_T \in \mc X }\{D_T\,X_T| d_L \leq D_T
\leq d_H\}\big)= {\rho_0}\big(d_L X_T^+ - d_H X_T^-\big), $ where
$v(x) =-( d_L x^+ \,- d_H\,x^-)$. \hfill{$\Box$}
\begin{remark}
\rm When $D_T$ varies between two random variables $D_L$ and $D_H$
in $\mc X$, $0 \leq D_L \leq D_T\leq D_H \leq 1$, the functional in
(\ref{example:csarm_ambiguous_discount}) is a cash sub-additive risk
measure $\mc R^{{\rho_0}}(X_T) ={\rho_0}(-V(X_T)),$ where $V$ is the
random function $V(\omega, x) =-( D_L(\omega) x^+ \,-
D_H(\omega)\,x^-)$, convex, decreasing with respect to $x$, $V_x \in
[-1, 0]$, for any given $\omega \in \Omega$, and $\mc
F_T$-measurable for any given $x \in \mathbb{R}$. Notice that when
$D_L=D_T$, $\mc R^{{\rho_0}}(X_T)=\rho_0(D_T X_T)$.
\end{remark}
Next example of cash sub-additive risk measure is not related to
risky/ambiguous discount factor but to insurance risks. 
Following \citeasnoun{jarrow_2002}, the put option premium with zero
strike price may be used as a possible measure of the firm
insolvency risk. The expected losses are discounted using
the risk free gross return $r \geq 1$.

%
\begin{corollary}\textsf{Put premium risk measure.}
The premium of a put option with strike price zero and maturity $T$,
\begin{equation}\label{FPPRM}
\mc R_p(X_T):=\frac 1 r \,
\mathbb{E}_{\mathbb{P}}\big[(-X_T)^+\big ],
\end{equation} is a coherent cash
sub-additive risk measure as a function of the underlying asset
price $X_T$.
\end{corollary}
\textit{Proof.} The cash sub-additive risk measure in
(\ref{example:csarm_ambiguous_discount})  coincides with the put
option premium $\mc R_p$ when $\rho_0(\cdot)= \mathbb
E_\mathbb{P}[-\, (\cdot)]$, $d_L=0$ and $d_H=1/r$. \hfill{$\Box$}
\begin{remark}\rm
For any given strike price $K$ the premium of a put option, $\mc
R_p(X_T):= \frac 1 r \, \mathbb{E}_\mathbb{P}\big[(K -X_T)^+\big
]$ is a cash sub-additive risk measure. This follows  setting in
equation (\ref{example:csarm_ambiguous_discount}) $\rho_0$ equals
to the non normalized risk measure $\rho_0(X_T)=
\mathbb{E_\mathbb{P}}[K-X_T]$ and $-v(x)=\frac1r \max (K-x, 0)$.
\end{remark}

\subsubsection{Composing cash additive risk measures and  convex
functions}\label{examples}
Generalizing the previous examples we show that $\rho_0(-V)$ is a
cash sub-additive risk measure, where $V$  is a random function $V:
\Omega \times \mathbb{R} \longrightarrow \mathbb{R}$, $V(\omega,
x)$, such that,
for any $\omega \in \Omega$,  $V(\omega, \cdot)$ is,
lower-semicontinuous (lsc), decreasing, convex  and $V(\omega,0)=0$,
$V_x\in [-1,0]$, and
for any $x \in \mathbb{R},$ $V(\cdot, x)$ is $\mc F_T$-measurable.
Moreover, $\rho_0(-V)$ can be represented in terms of finitely
additive measures  and $\mc F_T$-measurable ``discount factors" over
a set of possible scenarios that can be chosen according to the
beliefs of the agent/regulator.

From standard results in convex analysis $V(\omega, x)= \sup_{y
\in \mathbb{Q}}\{ x y -\beta_T(\omega, y) \}$, where $\beta_T $ is
the random convex Fenchel transform of $V$,
$\beta_T(\omega,y):=\sup_{x \in \mathbb{Q}}\{x y- V(\omega, x)\}$.
Notice that $\beta_T $ is finite only if $y \in [-1,0]$ as $V_x\geq
-1$.
For example, the Fenchel  transform of $v(x)= -( D_L x^+ \,-
D_H\,x^-)$ is $\beta_T(y)=l^{\mc D}(-y)$, where $l^{\mc D}$ is the
convex indicator function of the set $\mc D=[D_L, D_H ]$, equal to
$0$ on $\mc D$ and $\infty$ otherwise.
While $V_x \geq -1$ is a necessary condition to obtain a cash
sub-additive functional, the decreasing monotonicity ($V_x\leq 0$)
and convexity of $V$  insure the convexity and decreasing
monotonicity of $\rho_0(-V)$.
%
\begin{proposition}\label{prop:rho(-v)}
Let $V$ be a random, lsc, decreasing convex function as above and
$\beta_T$ the convex Fenchel transform of $ V$. Let $\rho_0 $ be a
cash additive risk measure defined on $\mc X$ with minimal penalty
function $\alpha_0$.
$\mc R^{\rho_0, V}(X_T):= \rho_0(-V(X_T))$ is a cash sub-additive
risk measure, derived from the spot risk measure $\rho_0$ by
assessing discount factors ambiguity through the penalty function
$\beta_T$,
\begin{equation}\label{dual_rho_v_firststep}
\mc R^{\rho_0, V}(X_T)= \sup_{D_T \in \mc X }\big\{
\rho_0\left(D_T\,X_T+ \beta_T(-D_T)\right)\,|\, 0 \leq D_T \leq
1\big\}.
\end{equation}
Moreover,  $\mc R^{\rho_0,V}(X_T)=\rho_0(-V(X_T))$ admits the dual
representation
\begin{eqnarray}\label{dual_rho_v}
&&\mc R^{{\rho_0}, V}(X_T)   =  \sup_{Q_0 \in \mc M_{1, f},
\,D_T\in \mc X}\big\{ \mathbb{E}_{Q_0 }[-D_T\,X_T  ] -
\alpha^{{\rho_0}, V}(Q_0 \,, D_T)|\, \, 0 \leq D_T\leq1\Big\}
\\&&  \alpha^{\rho_0,V} (Q_0 , D_T):=\>
\alpha_0(Q_0) + \mathbb{E}_{Q_0}[\beta_T(-
D_T)].\label{alpha-overlinerho-v}
\end{eqnarray}

\end{proposition}
For instance,  if $\rho_0$ is the coherent worst case risk measure,
that is, $\rho_0(X_T)={\rho}_{max}(X_T)= \sup_{Q_0 \in \mc
M_{1}}\mathbb{E}_{Q_0}[-X_T]$, then
$\mc R^{{\rho_0}, V}(X_T) = {\rho}_{max}(-V(X_T)) = \|
-V(X_T)\|_{\infty}= -V(\|X_T\|_{\infty})$ and $\alpha^{{\rho_0},
V}(Q_0, D_T):= \mathbb{E}_{Q_0}[\beta(-D_T)]$.
\begin{remark}\rm
Representation (\ref{dual_rho_v})--(\ref{alpha-overlinerho-v})
provides a better understanding of the different risks involved in
the evaluation of the risky position $X_T$. The scenarios could be
exogenously determined, for instance by some regulatory
institution. The penalty function $\alpha^{\rho_0,V}$ depending on
the ambiguous model and ambiguous discount factor could be
determined by the preferences of the economic agent on $Q_0$ and
$D_T $.
\end{remark}
%
\begin{remark}\textsf{Robust expected utilities and cash sub-additive risk measures.}
\rm By definition, the functional $\mc R^{\rho_0, V}$ admits a
representation in terms of the ambiguous model and the convex
function on the risky positions,
$\mc R^{\rho_0, V}(X_T)=\sup_{Q_0 \in \mc M_{1,
f}}\big\{\mathbb{E}_{Q_0 }[V(X_T ) ]- \alpha_0(Q_0)\big\}$.
The opposite of $\mc R^{\rho_0, V}$ can be viewed as examples of
robust expected utilities associated with concave functions $U=-V$
and concave penalty functions $\tilde \alpha_0 = - \alpha_0$,
that is $-\mc R^{\rho_0, V}(X_T)= \inf_{Q_0 \in \mc M_{1,
f}}\big\{\mathbb{E}_{Q_0 }[U(X_T)]- \tilde{\alpha_0}(Q_0) \big\}$.
Notice that $U$ does not satisfy the Inada conditions.
For robust expected utilities see, for instance,
\citeasnoun{schied_2004a} and
\citeasnoun{maccheroni_marinacci_Rustichini_2004}.
\end{remark}
\textit{Proof.} Firstly we prove that $\mc R^{\rho_0, V}$ is a cash
sub-additive risk measure. \textit{Decreasing monotonicity:} The
increasing monotonicity of $-V$ and the decreasing monotonicity of
$\rho_0$ imply the decreasing
monotonicity of $\mc R^{\rho_0, V}$.\\
\textit{Convexity:} The concavity of $-V$, the decreasing
monotonicity and the convexity of $\rho_0 $
imply the convexity of $\mc R^{\rho_0, V}$.\\
\textit{Cash sub-additivity:} $\mc R^{\rho_0, V}(X_T+m)+m
=\rho_0(-V(X_T+m1_T))+m= \rho_0(-V(X_T+m1_T)-m)$ is increasing in
$m$ if
$-V(X_T+m1_T)-m$ is decreasing in $m$. As $V_x>-1$ the result follows.\\
\textit{Representations:}  To prove (\ref{dual_rho_v_firststep})
we observe that
$$
\rho_0(-V(X_T))= {\rho_0}\big(\inf_{-1\leq y\leq 0}\{-X_T
y+\beta_T(y)\}\big)
 =\rho_0\big(\inf_{D_T\in \mc X}\{ D_T X_T +\beta_T(-D_T)\,| \,0
\leq D_T \leq 1 \}\big).
$$
From the decreasing monotonicity of $\rho_0$, for any
$\widetilde{D}_T \in \mc X,\, \, 0\leq \widetilde{D}_T\leq 1$ we
have
\begin{eqnarray}\label{allo}
\rho_0\big(\inf_{D_T\in \mc X}\{ D_T X_T+\beta_T(-D_T) \,| \,0 \leq
D_T \leq 1 \}\big) \geq \rho_0\left( \widetilde{D}_T X_T+
\beta_T(-\tilde{D}_T)\right).
\end{eqnarray}
The result follows setting $\widetilde{D}_T= D^*_T$ in equation
(\ref{allo}), where $D^*_T\in \mc X$ is  the element achieving the
\textbf{$\inf_{D_T\in \mc X}\{ D_T X_T+\beta_T(-D_T) \,| \,0 \leq
D_T \leq 1 \}$}.
Finally, representations
(\ref{dual_rho_v})--(\ref{alpha-overlinerho-v}) are obtained from
the dual representation of $\rho_0$ and from
(\ref{dual_rho_v_firststep}). \vspace{2mm}\hfill{$\Box$}\\
The penalty function $\alpha^{{\rho_0}, v}$ in
(\ref{alpha-overlinerho-v}) is not the minimal one.
As any pair $(Q_0, D_T)$ defines a unique additive set function
$\mu$ absolutely continuous with respect to $Q_0$, $d\mu:=D_T
dQ_0$, $0 \leq \mu(\Omega)\leq 1$, the functional $\mc R^{\rho_0,
V}$ can be rewritten as
\begin{equation*}
\mc R^{\rho_0, V}(X_T) = \sup_{\mu \in \mc M_{1,f}(\mc F_T)} \Big\{
\mu(-X_T) - \gamma(\mu)
 \,| \,0\leq \mu(\Omega)\leq 1 \Big\},
 \end{equation*}
where  $\mu(-X_T):=\int -X_T(\omega) \mu(d\omega) $ and $\gamma(\mu)
= \inf_{Q_0 \in \mc M_{1, f}}{\big\{ \alpha_0(Q_0) +
\mathbb{E}_{Q_0} \left[ \beta_T\left(-\frac{d\mu}{dQ_0}\right)
\right]} \big\}$ for any $\mu $ such that $d\mu= D_T dQ_0$, $0\leq
D_{0, T}\leq 1$, and $\gamma= \infty$ otherwise.

Next section gives the dual representation of the cash sub-additive
risk measures $\mc R$ in terms of the minimal penalty function.

\section{Minimal cash additive extension of $\mc R$ and duality}
\label{sec:cash_sub_additive_dual_reperesentation}
In this section we study the dual representation of cash
sub-additive risk measures.
To obtain duality results we can either use convex analysis tools
(for instance adapting the techniques for convex risk measures in
\citeasnoun{kratschmer_2007}), or recover  cash sub-additive risk
measures by enlarging the space of risky positions.
We adopt the second approach because of its richer financial
interpretation, despite the fact that the first one could be less
involved.
Our approach provides an interesting interpretation of cash additive
and cash sub-additive risk measures where default events or
stochastic num\'{e}raires are taken into account.
Taking a classical procedure in credit risk modeling, we consider a minimal enlargement of the sample space
$\Omega$ and we extend the cash sub-additive risk measure $\mc R$
into a cash additive risk measure, which is in a one to one
correspondence with $\mc R$. This allows to derive properties of
$\mc R$ and the dual representation using classical theory on
cash additive risk measures.

\citeasnoun{cheridito_kupper_2006} use a similar procedure to
decompose dynamic cash additive risk measures in one-step generators
and to provide a dual representation of these generators.
Interestingly, these generators are cash sub-additive risk measures
with opposite sign\footnote{We thank the referee for
highlighting this result.}.

While in the dual representation of cash additive risk measures the
set functions $Q$ are normalized to one in $ {\mathcal
M}_{1,f}({\mathcal F}_T)$,
the dual representation of cash sub-additive risk measures is based
on finite additive set functions $\mu$ with total mass
$0\leq\mu(\Omega)\leq1$, called sub-probability measures, whose
set is denoted by ${\mathcal M}_{s,f}({\mathcal F}_T)$.
A simple procedure to obtain a cash additive risk measure using a
cash sub-additive risk measure $\mc R$ is as follows. While $\mc
R$ is not cash additive with respect to $X_T \in \mc X$, the
bivariate function $\hat{\rho}(X_T, x):= \mc R(X_T1_T - x 1_T)- x$
as a function of the pair $(X_T,x)$ is cash additive.
In the sequel, we introduce the minimal measurable space where the
pair $(X_T,x)$ is the coordinate of a random variable and
$\hat{\rho}$ is a cash additive risk measure on these random
variables.
%
\subsection{Minimal extension of $\mc R$ into a cash additive risk measure}
\label{subsec:a minimal-extension-into-a-cash-additive}
Any pair $(X_T, x)$ where $X_T\in \mc X$ and $x \in \mathbb{R}$
can be viewed as the coordinates of a function defined on the
enlarged space $\widehat \Omega=\Omega\times\{0,1\}$ with element
$(\omega, \theta)$,
$$\widehat X_T(\omega,\theta):=
X_T(\omega)1_{\theta=1}+x 1_{\theta=0}.$$
$\widehat{\Omega}$ is endowed with the $\sigma$-algebra $\widehat{
\mc F}_T $ generated by the bounded random variables $\widehat X_T$.
Notice that $\widehat{ \mc F}_T $ is not the product
$\sigma$-algebra.
Let $\widehat{\mc X}$ be the linear space of all bounded random
variables $\widehat{X}_T$.
To denote $\widehat X_T \in\widehat{\mc X}$ we use its coordinates
$\widehat X_T=(X_T, x)$. The constant variables are denoted by
$\widehat{m}=(m,m)$ and $\widehat{m}= m1_{\theta=1}+ m
1_{\theta=0}=m$.
The event $\{\theta=0\}$ is atomic and all
$\widehat{\mc{F}}_T$-random variables are constant on this event.
The event $\{\theta=1\}$ models the risk affecting the num\'eraire
$1_T$. Intuitively, $\theta$ is associated with the default time
$\tau$ of the counterpart. The event $\{\theta=1\}$ is equivalent
to $\{\tau >T\}$.
The choice of the atomic $\sigma$-algebra $\mc {\widehat{F}}_T$
implies a one to one correspondence between  normalized additive
set function $\widehat{Q}$ in $\mc M_{1,f}(\mc {\widehat{F}}_T)$
and sub-probability set functions in $\mc M_{s,f}(\mc {F}_T)$ on
$(\Omega, \mc F_T)$.
Indeed,  any $\widehat{Q}$ in $\mc M_{1,f}(\mc {\widehat{F}}_T)$
can be decomposed as follows, $\forall \widehat X_T=(X_T, x) \in
\mc{ \widehat{X}}, $
\begin{equation}\label{hatQ}
\widehat{Q}(\widehat{X}_T)=
\widehat{Q}(X_T1_{\theta=1})+x\widehat{Q}(1_{\theta=0})=
\mu(X_T)+x(1-\mu(1)),
\end{equation}
where  $\mu(\cdot): = \widehat{Q}(\cdot1_{\theta=1})$ is an
additive sub-probability of $\mc M_{s,f}(\mc F_T)$.

The following proposition shows how to extend the cash sub-additive
functional $\mc R$ into a cash additive risk measure $\hat{\rho}$ on
$\mc {\widehat X}$ and the one to one correspondence.
\begin{proposition}\label{pro:hatrhofromR}
\it{ 1)} A normalized cash sub-additive risk measure $\mc R$ on $\mc
X$ defines a normalized cash additive risk measure $\widehat \rho$
on $\widehat{\mc X}$,
\begin{eqnarray}\label{hatrho}
\forall \ \widehat{X}_T = (X_T,x) \in \widehat{\mc X},\  \  \,\,
\hat{\rho}(\widehat{X}_T) :=\hat{\rho}\big((X_T, x)\big): = \mc
R\big(X_T - x1_T\big)- x.
\end{eqnarray}
Notice that $\hat{\rho}(X_T1_{\theta=1})=\mc R(X_T)$.\\
\it{2)} Any cash additive risk measure on $\widehat{\mc X}$
restricted to the event $\{\theta=1\}$ defines a cash sub-additive
risk measure which satisfies equation (\ref{hatrho}).
\end{proposition}
\begin{remark}{\rm
The cash sub-additive risk measure $\mc R$ can be used to measure
the risk of defaultable contingent claims $\widehat{X}_T$ when there
is no compensation ($x=0$) if the default occurs,~$\{\theta=0\}$.}
\end{remark}
The proof relies on the cash
sub-additive property to obtain a monotone decreasing functional.\\
\textit{Proof.} \textit{1) Cash additive:} Let $\widehat
X_T=(X_T,x)\in \widehat {\mc X}$ and $m \in \mathbb{R}$. By
definition,
$\hat{\rho}\big(X_T1_{\theta=1}+x1_{\theta=0}+ m1_{\theta=1} +
m1_{\theta=0}\big)= \mc {R}\big(X_T +m 1_T-
(x+m)1_T\big)-(x+m) =\hat{\rho}\big(\widehat X_T )-m.$\\
\textit{Decreasing monotonicity:} Let $\widehat X_T=(X_T, x)$ and
$\widehat Y_T=(Y_T, y)$ $ \in \widehat{\mc X}$ such that $\widehat
X_T\geq \widehat Y_T$, that is $X_T\geq Y_T$ and $x\geq y$. From the
cash sub-additivity and the decreasing monotonicity of $\mc R$ it
follows that
$\hat{\rho}(\widehat{X}_T)=\mc R(X_T - x 1_T)- x  \leq \mc R(X_T-
y1_T)-y \leq \mc R(Y_T- y1_T)-y =\hat{\rho}(\widehat{Y}_T).$\\
\textit{Convexity}: We use the notation in equation
(\ref{baricenter}). From the convexity of $\mc R$, $\mc R \left(
Bar[X_I] \right) \leq Bar \left[ \mc R(X)_I \right]$. This implies
that
$\hat{\rho}(Bar[\widehat X_I])= \mc R \left( Bar \big[X_I-x_I\big]
\right)- Bar[x_I]\leq Bar\left[ \mc R \big( X-x\big)_I\right]-
Bar[x_I]= Bar \left[ \hat{\rho}(\widehat X)_I \right]$,
which shows the convexity of $\hat{\rho}$.\\
\textit{2)} Let $\check{\rho}$ be a cash additive risk measure on
$\widehat{\mc X}$. We have to show that $\mc
R^{\check{\rho}}(X_T):=\check {\rho}(X_T1_{\theta=1})$ is a cash
sub-additive risk measure.  The decreasing monotonicity and
convexity  follow  from the definition. The cash sub-additive
property is verified observing that  $\mc R^{\hat{\rho}}(X_T+m1_T)+m
= \hat {\rho}( (X_T+m) 1_{\theta=1})+m = \hat {\rho}( X_T
1_{\theta=1}- m1_{\theta=0})$ is increasing in $m$. $\hfill{\Box}$

\subsection{Dual representation of cash sub-additive risk measures}
In the next proposition we use the one to one correspondence in
equation (\ref{hatrho}) between $\hat {\rho}$ and $\mc R$ to
characterize cash sub-additive risk measures. We show that the
minimal penalty function of $\mc R$ and the minimal penalty function
$\hat{\rho}$ coincide and are concentrated on the set of
sub-probability measures $\mc M_{s,f}(\mc F_T)$.
Moreover, under the additional assumption  of continuity from below
of $\mc R$ the dual representation in terms of $\sigma$-additive
sub-probability measures is obtained.

\begin{theorem}\label{dual representation}
(a) Any cash sub-additive risk measure $\mc R$ on $\mc X$ can be
represented in terms of finitely additive sub-probability
measures,
\begin{equation}\label{Rdual}
\forall X_T \in \mc X, \quad \mc R(X_T1_T)= \sup_{\mu \in \mc
 M_{s,f}(\mc F_T)}\left\{ \mu(-X_T)- \alpha^{\mc R}(\mu)\right\},
 \  \  \alpha^{\mc R}(\mu):=\hat{\alpha}(\widehat{Q}),
\end{equation}
where $\mu(\cdot)= \widehat{Q}(\cdot1_{\theta=1})$ and
$\hat{\alpha}$ is any penalty function representing $\hat{\rho}$.
In particular, if $\hat{\alpha}$ is the minimal penalty function
for $\hat{\rho}$ then $\alpha^{\mc R}$ is the minimal penalty
function for $\mc R$ and $ \alpha^{\mc R}(\mu)
=\sup_{X_T \in \mc X}\left\{ \mu(-X_T) -\mc{R}(X_T) \right\}$.\\
(b) When  $\mc R$ is a cash sub-additive risk measure continuous
from below any  penalty function $\beta$ representing $\mc R$
 is concentrated on the class $\mc M_s(\mc F_T )$ of
$\sigma$-additive sub-probability measures, i.e., $\beta(\mu)<\infty
\Rightarrow \mu \mbox{ is } \sigma\mbox{-additive}.$
\end{theorem}
\textit{Proof.} \textit{(a)}  From Proposition
\ref{pro:hatrhofromR},  $\mc R(X_T 1_T)=
\hat{\rho}(X_T1_{\theta=1})$. Equation (\ref{Rdual}) is implied by
the dual representation of $\hat{\rho}$ and the one to one
correspondence between $\widehat{Q}$ and $\mu$:
$\widehat{Q}(\cdot1_{\theta=1})=\mu(\cdot)$.
Let $\hat{\alpha}$ be the minimal penalty function of $\hat{\rho}$.
By definition of the minimal penalty function,
\begin{eqnarray}
\nonumber \hat{\alpha}(\widehat{Q})&=& \sup_{\widehat X_T \in
\widehat{ \mc X}_T} \big\{ \mathbb{E}_{\widehat{Q}}[-
X_T1_{\theta=1}-x1_{\theta=0}]-\hat{\rho}(\widehat X_T) \big\}\\&=&
\nonumber \sup_{\widehat{X}_T\in \widehat{\mc X}}\big\{
\mathbb{E}_{\widehat{Q}}[-(X_T-x)1_{\theta=1}] -x-\mc{R}(X_T-x1_T)
+x\big\}
\\&=&\label{alpha_on_Qhat} \sup_{X_T \in \mc X_T}\big\{
\mathbb{E}_{\widehat{Q}}[-(X_T)1_{\theta=1}] -\mc{R}(X_T) \big\},\ \
\  \widehat{Q}\in \mc M_{1,f}(\widehat{\mc F}_T).
\end{eqnarray}
As $\widehat{Q}(\cdot1_{\theta=1})=\mu(\cdot)$,  from equation
(\ref{alpha_on_Qhat}) we have
$\alpha^{\mc R}(\mu):=\hat{\alpha}(\widehat{Q})
 =  \sup_{X_T \in \mc X} \left\{ \mu( -X_T) -\mc{R}(X_T)
\right\},$ showing that $\alpha^{\mc R}$ is the minimal penalty
function for $\mc R$.
\\
{\it{(b)}} If $\mc R$ is continuous from below the cash additive
$\hat{\rho}$ is continuous from below as a function of $\widehat
X_T= (X_T,x)$.   Then from Theorem \ref{Theorem dual representation}
follows that the penalty function of $\hat{\rho}$ is concentrated on
the class $ \mc M_1 (\widehat{\mc F}_T)$. This implies that the
penalty function of $\mc R$ is concentrated on the set of
$\sigma$-additive sub-probability $\mc M_s(\mc F_T)$.\hfill{$\Box$}
\\
Next corollary shows a representation of $\mc R$ where the penalty
functional depends on constants $c\in [0,1]$ and probability
measures.
\citeasnoun{frittelli_scandolo_2005} provide examples of general
capital requirement with similar representations. It is interesting
to observe that, among these, the only capital requirement that
satisfies the property of cash sub-additivity is the one reflecting
the agent's temporal risk aversion, which is related to the
uncertainty in the num\'{e}raire. For more details see Section 6 in
\citeasnoun{frittelli_scandolo_2005}.
\begin{corollary}
Any  cash sub-additive risk measure $\mc R$  can be represented as
follows
\begin{equation}\label{representationwithconstants}
\forall X_T \in \mc X, \quad \mc R(X_T1_T)= \sup_{(c, Q_T) \in
[0,1]\times  \mc M_{1,f}(\mc F_T)}\left\{
c\,\mathbb{E}_{Q_T}(-X_T)- \alpha^{\mc R}(c\cdot Q_T)\right\}.
\end{equation}
When $\mc R$ is continuous from below,  the penalty function
$\alpha^{\mc R}(c\cdot Q_T)$ is concentrated on the set $[0,1]\times
\mc M_1(\mc F_T)$ where $\mc M_1(\mc F_T)$ is set of
$\sigma$-additive sub-probability.
If $\,\inf_{X_T\in \mc X}{\mc R(X_T)} = -\infty$, the constants $c$
in formula (\ref{representationwithconstants}) are strictly
positive.

\end{corollary}
\textit{Proof.} Equation (\ref{representationwithconstants}) follows
by normalizing the sub-probability in equation (\ref{Rdual}), more
precisely defining, for any $\mu \in \mc M_{s,f}$ such that
$\mu>0$,\,  $Q_T(\cdot)=\mu(\cdot)/c$ where $c:= \mu(\Omega)$.
If $\mu=0$ then  $c=0$ and $Q_T\in \mc M_{1,f}(\mc F_T)$ is not
uniquely identified.
The condition $\inf_{X_T\in \mc X}{\mc R(X_T)}= -\infty $ implies
$-\alpha^{\mc R}(0)= \inf_{X_T\in \mc X}{\mc R(X_T)}= -\infty$
excluding $\mu=0.$
\hfill{$\Box$}\vspace{2mm}\\
The following representation, suggested by an anonymous referee,
provides a characterization of cash sub-additive risk measures in
terms of ambiguous ``zero-coupon bond'' viewed as a deterministic
discount factor. The risky positions are assessed via a family of
forward convex risk measures (see Definition \ref{forwardrm})
parameterized by the deterministic discounted factors\footnote{We
thank the referee for this stimulating suggestion.}.
\begin{corollary}
Any cash sub-additive risk measure $\mc R$ such that $\inf_{X_T\in
\mc X}{\mc R(X_T)} = -\infty$ is a worst discounted forward risk
measures, that is
\begin{equation}\label{eq:forwarddisountedwc}
 {\mc R}(X_T1_T)= \sup_{c \in (0,1]} c \cdot  \rho_{T,c}(-X_T),
\end{equation}
where $(\rho_{T,c})$ is a family of forward cash additive risk
measures such that the functional $(X_T,c)\in {\mc X}\times
(0,1]\rightarrow c\cdot \rho_{T,c}(-X_T)$ is convex.
 \end{corollary}
\textit{Proof.} Since any functional on the right side of equation
(\ref{eq:forwarddisountedwc}) is a cash-sub additive, convex and
monotone functional, their supremum shares the same property.\\
Vice versa, given a cash sub-additive risk measure $\mc R$ and his
dual representation, as $\inf_{X_T\in \mc X}{\mc R(X_T)} = -\infty$,
we can define the forward risk measures,
$$\rho_{T,c}(-X_T)=\sup_{Q_T \in \mc M_{1,f}(\mc
F_T)}\left\{\mathbb{E}_{Q_T}(-X_T)- \frac{\alpha^{\mc R}(c \cdot
Q_T)}{c} \right\}.$$
By definition, the family $c\cdot\rho_{T,c}(-X_T)$ is convex in both
arguments $(c,X_T)$ and $\mc R$ can be rewritten as in equation
(\ref{eq:forwarddisountedwc}). $\hfill$$\Box$

\section{Other cash additive extensions of $\mc R$} \label{sec:cash_sub_additive_credit_risk}
In this section we provide a representations of a cash sub-additive
risk measures $\mc R$ where the ambiguous discount
factor/num\'{e}raire is explicitly modeled as random variables of
$\mc X$.
While for the cash sub-additive risk measures generated via convex
functions (in Section \ref{Examples}) these representations were
easily obtained, to derive similar results for a generic $\mc R$ new
assumptions and more involved techniques are required.
To achieve this goal, we apply the same procedure  as in Section
\ref{sec:cash_sub_additive_dual_reperesentation} and we extend $\mc
R$ to a larger space that contains $\mc X$.
In this case the extension is not unique and requires the
introduction of an auxiliary a priori cash additive risk measure.
Then, for the cash sub-additive risk measures generated via convex
functions we propose another extension on the same enlarged space
obtained through a conditional risk measure.

\subsection{Cash sub-additive risk measures and  ambiguous discounted factors}
To define a linear space which contains  $\mc X$, the
$\sigma$-algebra $\widehat{\mc F}_T$ defined in Section
\ref{sec:cash_sub_additive_dual_reperesentation} is replaced by the
product $\sigma$-algebra $\mc G_T$.
On ($\Omega \times \{ 0,1 \}, \mc G_T$) any bounded $\mc
G_T$-random variable $\widetilde{X}_T$ can be represented as
$\widetilde{X}_T(\omega, \theta) = X_T^1(\omega) 1_{\theta=1} +
X_T^0(\omega) 1_{\theta=0 }$ and $X_T^0, X_T^1 \in \mc X $.
Let  $\widetilde{\mc X}$ be the linear space of all the bounded $\mc
G_T$-random variables $\widetilde{X}_T$. We refer to
$\widetilde{X}_T$ using the short notation $\widetilde{X}_T=(X_T^1,
X_T^0)$.
The diagonal elements $\widetilde{X}_T=(X_T, X_T)$ coincide with
$X_T$ and the corresponding $\sigma$-algebra with $\mc F_T$. This
identification was not possible for the random variables
$\widehat{X}=(X_T, x)$ defined in the previous section.

Now we discuss the probabilistic structure of ($\Omega \times \{
0,1 \}, \mc G_T$). Notice that in this section we consider
probability measures and not finite additive set functions.
\begin{definition}\label{QandD_T}\rm
For any given probability measure $\widetilde{\mathbb{ Q}}\in \mc
M_1(\mc G_T)$ let $\mathbb{Q}$ denote the restriction of
$\widetilde{\mathbb{Q}}$ to $\mc F_T$, $\mathbb{Q}:=
\widetilde{\mathbb{Q}}|\mc F_T$,  and $D_T\in [0,1]$ the $\mc
F_T$-conditional probability of the event $\{\theta=1\}$,
$\,D_T:={\mathbb{E}}_{\widetilde{\mathbb{ Q}}}[1_{\theta=1}|\mc
F_T]$, also called discount factor.
We denote $\overline{\mathbb{Q}}$ the probability measure associated
with the restriction of $\widetilde{\mathbb{Q}}$ to the event
$\{\theta=0\}$, which is uniquely determined by $(\mathbb{Q},D_T)$
\begin{equation}\label{Q}
\mathbb{Q}(X_T)=
\mathbb{Q}(D_TX_T)+(1-\mathbb{Q}(D_T))\overline{\mathbb{Q}}(X_T).
\end{equation}
$\overline{\mathbb{Q}}$ is a probability measure absolutely
continuous with respect to $\mathbb{Q}$, with Radon-Nikodym density
given by
$\overline{\Delta}_T:=\frac{d\overline{\mathbb{Q}}}{d\mathbb{Q}}=
\frac{1- D_T}{(1-\mathbb{Q}(D_T))}$, $0\leq \overline{\Delta}_T\leq
1,$ $\mathbb{Q}(\overline{\Delta}_T)=1$.
\end{definition}
For any $\widetilde{X}_T=X^1_T1_{\theta=1}+X^0_T1_{\theta=0}\in
\widetilde{\mc X}$,
\begin{equation}\label{tildeQ}
\widetilde{\mathbb{Q}}(X^1_T1_{\theta=1}+X^0_T1_{\theta=0})=
\mathbb{Q}(X_T^1D_T)+\mathbb{Q}(X_T^0(1-D_T))
=\mathbb{Q}(X_T^1D_T)+(1-\mathbb{Q}(D_T))\overline{\mathbb{Q}}(X_T^0).
\end{equation}
\begin{remark}\textsf{The interpretation of $D_T$ in  credit risk.}
{\rm In credit risk, $\theta$ is associated with the default time of
the counterpart $\tau$, where $\tau$  is a positive random variable
non $\mc F_T$-measurable. The event $\{\theta=1\}$ can be viewed as
$\{\tau>T\}$ and ${\mathbb{E}}_{\mathbb{\widetilde
Q}}[1_{\theta=1}|\mc F_T]$ as the conditional survival probability
function of $\tau$ at time $T$. $\widetilde X_T= X^1_T 1_{\theta=1}
+ X^0_T 1_{\theta=0}\in \widetilde{\mc X}$ is a defaultable
contingent claim that pays $X^1_T$ (at time $T$) if there is no
default ($\tau>T$) and $X^0_T$ otherwise.}
\end{remark}
In the sequel we extend $\mc R$ into a cash additive risk measure
$\tilde \rho$ on the enlarged space $\widetilde{\mc X}$. Via the
penalty function of $\tilde{\rho}$, a representation of cash
sub-additive risk measures will be given in terms of the ambiguous
probability measure and the ambiguous discount factor, both on the
original space of definition of $\mc R$.
To define this cash additive risk measure on $\widetilde{\mc X}$ we
use, as in Section \ref{subsec:a
minimal-extension-into-a-cash-additive}, the cash additive risk
measure $\hat{\rho}$ in (\ref{hatrho}).
In this case $\widetilde X_T=(X_T^1, X_T^0)\in \widetilde{\mc X}$
has two risky components and we introduce an a priori risk measure
$\overline{\rho}$ assessing the risk of the second component.
\begin{definition}\label{def:tilde-rho}
\rm Let $\mc R$ be a cash sub-additive risk measure and
$\overline{\rho}$ a cash additive risk measure both normalized and
defined on $\mc X$.
The functional on $\widetilde{\mc X}$
\begin{equation}\label{tilderho}
\tilde \rho(\widetilde{X}_T)= \tilde \rho(X_T^1, X_T^0):=\mc R(X_T^1
+ \overline{\rho}(X_T^0)1_T)+ \overline{\rho}(X_T^0)={\hat
\rho}\big(X_T^1,-\overline{\rho}(X_T^0)\big)
\end{equation}
and its restriction on $\mc X$,
\begin{equation}\label{rho-R-overline-rho}
\rho_{\mc R, \overline{\rho}}(X_T):= \mc R(X_T +
\overline{\rho}(X_T)1_T)+ \overline{\rho}(X_T)={\hat
\rho}\big(X_T,-\overline{\rho}(X_T)\big),
\end{equation}
are cash additive risk measures. Moreover, $\tilde{\rho}(X_T
1_{\theta=1})=\mc R(X_T 1_T)$.
\end{definition}
The following theorem shows that $\mc R$ can be written as a
function of probability measures  $\mathbb{Q}\in \mc M_1(\mc F_T)$
and $\mc F_T$-measurable discount factors $D_T \in \mc X$ using the
minimal penalty function of the cash additive risk measure
$\tilde{\rho}$. This representation is similar to the dual
representation (see equations
(\ref{dual_rho_v})--(\ref{alpha-overlinerho-v})) of cash
sub-additive risk measures generated by convex functions.

We consider penalty functions concentrated on the class of
probabilities measures assuming that $\mc R$ and $\overline{\rho}$
are continuous from below. This implies that also $\tilde {\rho}$
and $\rho_{\mc R, \overline{\rho}}$  are continuous from below.
\begin{theorem}\label{prop:tilderho}
Assume that the convex functionals $\mc R$ and $\overline{\rho}$ are
continuous from below. Let $\alpha^{\mc R} $ and $\overline{\alpha}$
be the minimal penalty functions of  $\mc R$ and $\overline{\rho}$,
respectively.
Let  $\tilde{\alpha}$ be the minimal penalty function of
$\tilde{\rho}$ defined in equation (\ref{tilderho}).
For any $\widetilde{\mathbb{Q}}\in \mc M_1(\mc G_T)$, let
$\mathbb{Q}$, $D_T$ and $\overline{\mathbb{Q}}$ be as in Definition
\ref{QandD_T}, such that
$\displaystyle{\frac{d\overline{\mathbb{Q}}}{d\mathbb{Q}}=\overline{\Delta}_T=\frac{1-
D_T}{(1-\mathbb{Q}(D_T))}}$.
\\
\it{1)} The cash sub-additive risk measure $\mc R$ can be
represented as
\begin{equation}
\label{RandAlphatilde} \mc R (X_T 1_T) = \tilde {\rho}(X_T 1_{\theta=1})
= \sup_{\mathbb{Q} \in \mc M_1(\mc F_T), \,D_T \in [0,1]} \big
\{\mathbb{E}_{\mathbb{Q}}(-{D_T}X_T)-
{\tilde{\alpha}}\big({D_T},\mathbb{Q}\big) \big\},
\end{equation}
where the minimal penalty $\tilde{\alpha}$ has the following form
\begin{equation}\label{alphatilde}
{\tilde{\alpha}}(\mathbb{\widetilde{Q}})=
{\tilde{\alpha}}(\mathbb{Q}, D_T) = \alpha^{\mc
R}(D_T\cdot\mathbb{Q})
+\left(1-\mathbb{Q}(D_T)\right)\overline{\alpha}(\overline{\mathbb{Q}}),\,
\  \   \widetilde{\mathbb{ Q}}\in \mc M_1(\mc G_T).
\end{equation}
Notice that $\mathbb{\widetilde{Q}} \in \mc Dom({\tilde{\alpha}})$
if and only if $\mathbb{Q}\cdot D_T\in  \mc Dom({\alpha}^{\mc R})$
and $\overline{\mathbb{Q}}\in \mc Dom(\overline {\alpha})$.
\\
\it{2)} The minimal penalty function of $\rho_{\mc R,
\overline{\rho}}$ in equation (\ref{rho-R-overline-rho}) is given by, for any $\mathbb{Q}\in \mc M_1(\mc F_T)$,
\begin{equation}\label{alpharho-R-overline-rho}
{\alpha}_{\mc R, \overline{\rho}}(\mathbb{Q})=  \inf_{
D_T,\overline{\mathbb{Q}}} \left\{\alpha^{\mc R}(D_T \cdot
\mathbb{Q})+\left(1-\mathbb{Q}(D_T)\right)
\overline{\alpha}(\overline{\mathbb{Q}})\,  \mid\,
\mathbb{Q}(\cdot)=\mathbb{Q}(D_T\cdot) +
(1-\mathbb{Q}(D_T))\overline{\mathbb{Q}}(\cdot) \right\}. \,
\end{equation}
\end{theorem}
\begin{remark}\rm
When  $\mc R$ and $\overline{\rho}$ are both coherent risk measures,
equation (\ref{RandAlphatilde}) reduces to
\begin{equation}
\mc {R}(X_T)=\tilde \rho({X}_T1_{\theta=1}) =\sup_{\mathbb{Q}\in \mc
M_1(\mc F_T), \,D_T \in [0,1]}
\big\{\mathbb{E}_{\mathbb{Q}}(-{D_T}X_T) \> |\>D_T\cdot\mathbb{Q}\in
\mc Dom(\alpha^{\mc R}), \> \overline{\Delta}_T\cdot \mathbb{Q} \in
\mc Dom(\overline{\alpha})\big\}.\nonumber
\end{equation}
\end{remark}
\textit{Proof.} {\textit 1) The representation
(\ref{RandAlphatilde}) of $\mc R$ follows from $\mc {R}(X_T
1_T)=\tilde \rho(X_T1_{\theta=1})$ and equation (\ref{tildeQ}). To
obtain the decomposition of the minimal penalty function in equation
(\ref{alphatilde}) we use the the representation of
$\widetilde{\mathbb{Q}}$ in terms of $\mathbb{Q} (D_T\cdot)$ and
$\overline{\mathbb{Q}}$ given in definition \ref{QandD_T}. From the
definition of $\tilde{\rho}$ and of the minimal penalty function we
have
\begin{eqnarray*}
\tilde{\alpha}(\widetilde{\mathbb{Q}})&=& \sup_{(X_T^1, X_T^0)\in
{\widetilde{\mc X}}} \left\{ \widetilde{\mathbb{Q}}(-X_T^1
1_{\theta=1} -X_T^0 1_{\theta=0})-\mc R(X_T^1+
\overline{\rho}(X_T^0)1_T)- \overline{\rho}(X_T^0)
 \right\}\\
 &=&
\sup_{(X_T^1, X_T^0)\in {\tilde{\mc X}}} \left\{
\widetilde{\mathbb{Q}}(-(X_T^1+\overline{\rho}(X_T^0))
1_{\theta=1})-\mc R(X_T^1+ \overline{\rho}(X_T^0)1_T)+
\widetilde{\mathbb{Q}}(-(X_T^0+\overline{\rho}(X_T^0)) 1_{\theta=0})
 \right\}.
\end{eqnarray*}
Using the change of variable $Y_T:=X_T^1+\overline{\rho}(X_T^0)$ and
equations (\ref{Q})--(\ref{tildeQ}) give the result
\begin{eqnarray*}
&&\tilde{\alpha}(\widetilde{\mathbb{Q}}) = \sup_{(Y_T, X_T^0)}
\left\{{\mathbb{Q}}(-Y_T\,D_T)-\mc R(Y_T)+
(1-\mathbb{Q}(D_T))\big[\overline{\mathbb{Q}}(-(X_T^0+\overline{\rho}(X_T^0))\big]
 \right\}\\
&&=\alpha^{\mc R}(D_T\cdot\mathbb{Q})+ (1-\mathbb{Q}(D_T))\sup_{ X
\in \mc A^{\overline{\rho}}} \left\{{\mathbb{Q}}(-X
\overline{\Delta}_T) )
 \right\}\\&&=
 \alpha^{\mc R}(D_T\cdot\mathbb{Q})+
(1-\mathbb{Q}(D_T))\,
\overline{\alpha}(\overline{\Delta}_T\cdot\mathbb{Q}).
\end{eqnarray*}
\textit{2)}  To obtain the penalty function ${\alpha}_{\mc R,
\overline{\rho}}$ of $\rho_{\mc R, \rho_0}$ we restrict
$\tilde{\rho}$ on  $\mc F_T$ and we use equation (\ref{Q})
\begin{eqnarray*}
\rho_{\mc R, \rho_0}(X_T) & = & \sup_{\mathbb{Q}\in \mc M_{1}(\mc
F_T)} \left\{ \mathbb{Q}(-X_TD_T) +
(1-\mathbb{Q}(D_T))\overline{\mathbb{Q}}(-X_T)\right.
\\&&\  \  \  \  \  \  \  \  \  \  \   \quad\quad\quad\quad
\left. -
\big(\alpha^{\mc R}(D_T \cdot \mathbb{Q})+ (1-\mathbb{Q}(D_T)
\overline{\alpha}(\overline{\mathbb{Q}}) \big) \right\}
\\
&=& \sup_{\mathbb{Q}\in \mc M_{1}(\mc F_T)} \left\{
\mathbb{Q}(-X_T)- \big(\alpha^{\mc R}(D_T \cdot \mathbb{Q})+
(1-\mathbb{Q}(D_T)) \overline{\alpha}(\overline{\mathbb{Q}})\big)
\right\}.
\end{eqnarray*}
Observing that for a given $\mathbb{Q} \in \mc M_{1}(\mc F_T)$ more
then one pair $(D_T, \overline{{\mathbb Q}})$, $D_T \in \mc X$, $D_T
\in [0,1]$, can verify $\mathbb{Q}(-X_T D_T) + (1-\mathbb{Q}(D_T))
\overline{\mathbb Q}(-X_T))= \mathbb{Q}(X_T)$ yields the equation
(\ref{alpharho-R-overline-rho}).
Similar calculations show that ${\alpha}_{\mc R, \overline{\rho}}$
is the minimal penalty function. \hfill{$\square$}

\subsection{Conditional risk measures and extensions on  $\widetilde{ \mc X}$}
This section reinterprets the cash sub-additive risk measures $\mc
R^{\rho, V} = \rho(-V)$ studied in Section~\ref{examples}. These
risk measures are now represented as the composition of the
unconditional cash additive risk measure $\rho$ and the
\emph{conditional} cash additive risk measure generated by the
random function $V$.
We obtain the result introducing a more natural extension of $\mc
R^{\rho, V}$ called $\check {\rho}^V$ to the enlarged space
$\widetilde {\mc X}$.
The restriction of $\check{\rho}^V$ to the space $\mc X$ is $\rho$
itself,  and $\check {\rho}^V$ can be obtained composing $\rho $
with a cash additive conditional risk measures.
Moreover, we show that any cash additive risk measure on
$\widetilde{ \mc X}$ generated from $\rho$ via a conditional cash
additive risk measure is associated  to a cash sub-additive risk
measure generated by a convex function.

As in Section \ref{examples}, in the sequel $\rho$ denotes a
normalized cash additive risk measure and  $V(\omega, x)$ an $\mc
F_T$-measurable random functional convex monotone decreasing such
that $V(0)=0$ and with left derivative $V_x\in [-1,0]$. From
Proposition \ref{prop:rho(-v)} we know that $\mc R^{\rho, V}(X_T):=
\rho(-V(X_T))$ is a cash sub-additive risk measure on $\mc X$.

\begin{proposition}\label{pro:widetildeRand_Rho}
On the enlarged space $\widetilde{\mc X}$ any cash additive risk
measure $\rho$ and any random function $V$  define a cash additive
risk measure,
\begin{equation}\label{check_Rand_Rho}
\check{\rho}^V(X_T^1 1_{\theta=1} +X_T^01_{\theta=0}):= \rho
\big(-V(X_T^1-X_T^0) + X^0_T\big), \ \ X_T^1 1_{\theta=1}
+X_T^0\,1_{\theta=0}\in \widetilde{\mc X}.
\end{equation}
$\check{\rho}^V$  coincides with $\mc R^{\rho, V}$ on $\{\theta=1\}$
and with $\rho$ on  $\mc X\subset \widetilde{\mc X} $:
\begin{equation*}
\check{\rho}^V(X_T1_{\theta=1})=\rho \big(-V(X_T)\big)=\mc R^{\rho,
V}(X_T)\  \mbox{\  and \ }\   \check{\rho}^V((X_T, X_T))=\rho\big(
X_T\big).
\end{equation*}
\end{proposition}
Requiring $V$ decreasing monotone and such that $V_x\in [-1,0]$ is
crucial to obtain $\check{\rho}^V$ decreasing monotone (see proof
below).
\\
\textit{Proof.} \textit{Decreasing monotonicity:} $\check{\rho}^V$
is decreasing monotone if  $V(X_T^1-X_T^0) - X_0$ is decreasing
monotone with respect to $(X^1_T, X^0_T)$. Let
$\widetilde{X}_T=(X_T^1, X_T^0)\geq \widetilde{Y}_T=(Y_T^1, Y_T^0)$,
that is $X_T^1 \geq Y_T^1$ and $X_T^0 \geq Y_T^0$.
As $V(x+m)+m$ is not decreasing in $m$, $V(X_T^1 -X_T^0)-X_T^0$ is
not increasing in $X_T^0$, then $V(X_T^1 -X_T^0)-X_T^0 \leq V( X_T^1
-Y_T^0)-Y_T^0 \leq V(Y_T^1 -Y_T^0)-Y_T^0,$ where the last inequality
is due to the decreasing monotonicity of $V$.\\ \textit{Cash
additivity} and \textit{convexity} follow from the definition of
$\check{\rho}^V$.
$\hfill{\Box}$\\
Now we recall the definition of conditional risk measures that in
our setting\footnote{For conditional risk measures see
\citeasnoun{bion_nadal_2004}, \citeasnoun{detlefsen_scandolo_2005}
and references therein.} reads as follows.
\begin{definition}\rm
1) A {\it cash additive conditional risk measure on $\mc F_T$} is a
monotone decreasing convex functional, $\tilde{\rho}_{\mc F_T} :
\widetilde{ \mc{X}}\rightarrow \mc{X}$ which satisfies  the  $\mc
F_T$-cash additive axiom, that is \\ $\forall \widetilde{X} \in
\widetilde{\mc X}, \, \forall Y \in \mc X, \quad \tilde{\rho}_{\mc
F_T}( \widetilde X + Y)
=\tilde{\rho}_{\mc F_T}(\widetilde X) -Y$.\\
2) $\tilde{\rho}_{\mc F_T}$ is \textit{regular} if for any $F_T \in
\mc F_T$, $ \widetilde  X_T \in \widetilde{\mc X},\,\,
\tilde{\rho}_{\mc
F_T}(1_{F_T} \widetilde X_T)=1_{F_T}\tilde{\rho}_{\mc F_T}(\widetilde X_T)$.\\
3) A cash additive risk measure $\check{\rho}$ on
$\widetilde{\mc{X}}$ is \textit{generated from $\rho$ via a
conditional risk measure} if there exists a cash additive
conditional risk measure on  $\mc F_T$, $\tilde{\rho}_{\mc F_T}$
such that,
$\check{\rho}(X_T^1, X_T^0)= \rho( -\tilde{\rho}_{\mc
F_T}((X_T^1,X_T^0)).$
\end{definition}
It easy to see that any conditional risk measure on $\mc F_T$ is
completely determined by its value on the set $\{\theta=1\}$. This
observation leads to the following proposition.
\begin{proposition}\label{pro:conditionalRM}
Any $\mc F_T$-measurable random function $V$ defines a cash additive
conditional risk measure on $\mc F_T$, $\tilde{\rho}^{V}_{\mc F_T} :
\widetilde{ \mc{X}}\rightarrow \mc{X}$, given by
\begin{equation}\label{v-conditional-rm-relation}
\tilde{\rho}^{V}_{\mc F_T}(X_T 1_{\theta=1}):= V(X_T)\  \mbox{ \  or
equivalently by \ }\ \tilde{\rho}^{V}_{\mc F_T}((X_T^1,X_T^0)):=
V(X_T^1 -X_T^0)-X_T^0.
\end{equation}
Conversely, any regular and continuous from above cash additive
conditional risk measure on $\mc F_T$, $\tilde{\rho}_{\mc F_T} :
\widetilde{ \mc{X}}\rightarrow \mc{X}$, generates a convex random function 
$\widetilde{V}^{\mc F_T}(\lambda):= \tilde{\rho}_{\mc
F_T}(\lambda1_{\theta=1})$ which
satisfies~(\ref{v-conditional-rm-relation}).
\end{proposition}
\textit{Proof.} \textit{Decreasing monotonicity:} We refer the
reader to the proof of decreasing monotonicity in
Proposition~\ref{pro:widetildeRand_Rho}.  \textit{$\mc F_T$-cash
invariance} and
 \textit{convexity}  follow respectively from the definition of
$\tilde{\rho}^{V}_{\mc F_T}$
and the convexity of $V$.\\
\textit{Conversely:} Define  $\widetilde{V}^{\mc F_T}(\omega
,\lambda):= \tilde{\rho}_{\mc F_T}(\lambda1_{\theta=1}(\omega))$.
$V(\omega ,\lambda)$ is $\mc F_T$-measurable convex and monotone
decreasing functional such that $\widetilde{V}^{\mc F_T}(0)=0$ and
$\widetilde{V}^{\mc F_T} \in [-1,0]$.
For the regularity of $\tilde{\rho}_{\mc F_T}$  the previous
definition can be extended to all the simple $\mc F_T$-random
variables $\sum \lambda_i 1_{A_i}$, where the sets $A_i\in \mc F_T$
and $\{A_i\}_{i=1,\ldots,n}$ form a partition of $\Omega$. Hence
$\tilde{\rho}_{\mc F_T}(\sum \lambda_i 1_{A_i})=\sum 1_{A_i}
\widetilde{V}^{\mc F_T}(\lambda_i)$.
The continuity from above of $\tilde{\rho}_{\mc F_T}$ allows to
extend the definition  to positive $X_T\in \mc X$ and then to any
arbitrary $X_T\in \mc X$ using standard analysis tools.
$\hfill{\Box}$

The following theorem states the main result of this section showing
that any cash sub-additive risk measure of the form $\mc R^{\rho,
V}= \rho(-V)$ can be extended into a cash additive risk measure
which is generated from $\rho$ via a conditional risk measure.
Conversely, any cash additive risk measure $\check{\rho}$ on
$\widetilde{\mc{X}}$ generated from $\rho$ via a conditional risk
measure is associated to a cash sub-additive risk measure of type
$\mc R^{\rho, \check{V}_{\mc F_T}}$.
%
%
\begin{theorem}\label{cashsubadditive_conditional}
The cash additive risk measure $\check{\rho}^{V}$ in equation
(\ref{check_Rand_Rho}) is generated from $\rho$ via the conditional
risk measure $\tilde{\rho}^{V}_{\mc F_T}$  in
(\ref{v-conditional-rm-relation}) associated with $V$, that is
\begin{equation}
 \check{\rho}^V(X_T^11_{\theta=1}+X_T^01_{\theta=0})= \rho
\big(-V(X_T^1-X_T^0) + X_0\big)=\rho\left(-\tilde{\rho}^{V}_{\mc
F_T}(X_T^11_{\theta=1}+X_T^01_{\theta=0}) \right).
\end{equation}
Moreover,
\begin{equation}\label{ciaociaociao}
\mc R^{\rho, V}(X_T)=
\check{\rho}^V(X_T1_{\theta=1})=\rho\left(-\tilde{\rho}^{V}_{\mc
F_T}(X_T 1_{\theta=1})\right).
\end{equation}
Conversely, to any cash additive risk measure $\check{\rho}(\cdot)=
\rho( -\tilde{\rho}_{\mc F_T}(\cdot))$ on $ \widetilde {\mc X}$
generated by a cash additive conditional risk measure
$\tilde{\rho}_{\mc F_T}$ on $\mc F_T$ is associated a cash
sub-additive risk measure of the following form $\mc R^{\rho,
{\check{V}}^{\mc F_T}}(X_T)=\rho\big(-{\check{V}}^{\mc
F_T}(X_T)\big)$ where ${\check{V}}^{\mc F_T}(X_T)= \tilde{\rho}_{\mc
F_T}(X_T 1_{\theta=1})$.
\end{theorem}
\textit{Proof.} The proof follows easily from the previous
considerations.\hfill{$\Box$}\\
Equation (\ref{ciaociaociao}}) suggests that the risk of the future
position $X_T$ depends on the risk/ambiguity on the underlying asset
model (the unconditional risk measure $\rho $) and on the
risk/ambiguity on interest rates (the conditional risk measure
$\tilde{\rho}_{\mc F_T}$) or more in general on the risk affecting
the num\'{e}raire.

\section{Optimal derivative design and inf-convolution}\label{sec:Optimal derivative design and inf-convolution}
The problem of designing the optimal transaction between two
economic agents has been largely investigated both in the insurance
and in the financial literature.
The risk transfer between the agents takes place through the
exchange of a derivative contract and the optimal transaction is
determined by a choice criterion.
For example, in \citeasnoun{barrieu_elkaroui_2006} the choice
criterion is given by the minimization of the risk of the agent's
exposure and the risk is assessed using forward cash additive risk
measures.
Using cash sub-additive risk measures we study  this problem in a
general framework that allows for ambiguous discount rates.
We focus on the problem of the risk transfer between two agents who
determine today the reserve to hedge the future exposure when the
discount factor for the maturity of interest is ambiguous.
To account for this ambiguity the agents collect the reserve using
cash sub-additive risk measures and the decision criterium is the
minimization of their reserves.

\subsection{Transaction feasibility and optimization program}
Let $A$ and $B$ be the two agents  and suppose that they are
evolving in a uncertain universe modeled by the probability space
$\left(\Omega ,\mc F_T\right )$.
Agent $A$ is exposed towards a non-tradable risk that will impact
her wealth $X_T^A\in \mc {X}$ at the future date $T$.
To reduce her risk exposure and the reserve associated, $A$ aims at
issuing  a derivative contract $H_T\in \mc {X}$ with maturity $T$
and selling it to the agent $B$  for a price $\pi_0$.
Agent $B$ will enter the transaction only if this transaction
reduces or leaves unchanged the reserve that she has to put aside to
hedge her future exposure $X_T^B\in \mc {X}$.
The objective is to find the optimal structure $\left(H_T,\pi_0
\right)$ according to the decision criterion of the agents given by
their cash sub-additive risk measure $\mc R_{A}$ and $\mc R_{B}$.

If the agents agree on the transaction, at time zero $B$  pays
$\pi_0$ to $A$. At time $T$ the terminal wealths of the agents $A$
and $B$ are $X_T^A-H_T$ and $X_T^B+H_T$, respectively.
$A$  aims at minimizing the current reserve $\mc R_{A}\left(
X_T^A-H_T \right)$ for the future exposure $X_T^A-H_T$, knowing that
today she receives $\pi_0$ from $B$,
\begin{eqnarray}\label{minprogram}
\inf_{H_T \in \mc{X},\pi_0 }\mc R_{A}\left( X_T^A-H_T \right)-\pi_0.
\end{eqnarray}
The constraint to the optimization program (\ref{minprogram}) is
that $B$ enters the transaction. This happens when buying $H_T$ for
$\pi_0$ reduces or leaves unchanged the reserve $\mc R_{B}\left(
X_T^B\right)$ that $B$ would collect not entering the transaction,
\begin{eqnarray}\label{ciaociao}
\mc R_{B}\left(X_T^B+ H_T \right)+ \pi_0 \leq \mc R_{B}\left(
X_T^B\right).
\end{eqnarray}
The pricing rule of the $H_T$-structure is fully determined by the
buyer $B$ simply binding the constraint at the optimum in equation
(\ref{ciaociao}),
\[
\pi_0 ^{*}=\pi_0 ^{*}\left(H_T\right) =\mc R_{B}\left( X_T^B \right)
-\mc R_{B}\left( X_T^B+ H_T \right).
\]
This price  $\pi_0 ^{*}$ corresponds to an ``indifference'' pricing
rule from the point of view of the agent $B$ as $\pi_0 ^{*}$ gives
the maximum amount that agent $B$ is ready to pay to enter the
transaction. Given $\pi_0 ^{*}$, the optimization program in
(\ref{minprogram}) becomes
\begin{equation}\label{eq:optimization_program_1}
{\mc R}_{A,B}(X_T^A, X_T^B) :=\inf_{H_T \in \mc{X}}\mc R_{A}\left(
X_T^A-H_T \right)+\mc R_{B}\left( X_T^B+ H_T \right),
\end{equation}
where the optimal transaction $H_T^*$ attains the infimum.

\subsection{Optimal transaction and inf-convolution}

The risk transfer problem in equation
(\ref{eq:optimization_program_1}) can be rewritten as an
inf-convolution of cash sub-additive risk measures on $\mc X$.
Indeed  defining $F_T:= X_T^B+ H_T\in \mc X$ we have
\begin{equation}\label{inf_conv_csarm}
{\mc R}_{A,B}(X_T^A,   X_T^B) =\inf_{ F_T \in \mc X }
 \big\{ {\mc R}_A(X_T^A+ X_T^B - F_T)
+ {\mc R}_{B}(F_T) \big\}=: {\mathcal R}_{A}\square {\mathcal
R}_{B}(X_T^A+ X_T^B),
\end{equation}
where $\square$ denotes the inf-convolution.
The value of ${\mc R}_{A,B}(X_T^A,   X_T^B)$ can be interpreted as
the residual  measure of risk after the transaction $F_T$ has
occurred. This residual  measure of risk depends on the initial
exposures $X_T^A$ and $X_T^B$. The transaction induces an optimal
redistribution of the  risks of the agents.
In the following we show that  ${\mathcal R}_{A}\square {\mathcal
R}_{B} $ is a cash sub-additive risk measure completely
characterized  by $\mc R^A$ and $\mc R^B$ and we provide its dual
representation.
Also in this case, instead of using convex analysis tools to prove
these results we exploit the one to one correspondence between $\mc
R$ and the cash additive risk measure $\hat{\rho}(\widehat{X}_T)=
\mc R \big(X_T - x1_T\big)- x$ defined on $\widehat{\mc X}$  and
given in equation (\ref{hatrho}).
We show that the inf-convolution of cash sub-additive risk measures
on $\mc X$ is equal to the inf-convolution of their corresponding
cash additive risk measures $\hat{\rho}$ on $\widehat{\mc X}$.
\begin{lemma}
The inf-convolution of ${\mathcal R}_{A}$ and ${\mathcal R}_{B}$ on
$\mc X$ in equation (\ref{inf_conv_csarm}) corresponds to the
inf-convolution of the cash additive extensions of $\mc R_A$ and
$\mc R_B$ on $\widehat {\mc X}$,
\begin{equation}\label{inf-conv-cashsubadd-cashadd}
{\mathcal R}_{A}\square {\mathcal R}_{B}(X_T^A+ X_T^B) =
\hat{\rho}_{A}\square \hat{\rho}_{B}(\widehat X_T^A+ \widehat
X_T^B), \mbox{ where } \widehat X_T^A := X_T^A 1_{\theta=1},\
\widehat X_T^B: = X_T^B1_{\theta=1}.
\end{equation}
${\mathcal R}_{A}\square {\mathcal R}_{B}(X_T^A+ X_T^B)$ is the
infimum on $F_T\in \mc X$, while $\hat{\rho}_{A} \square
\hat{\rho}_{B}(\widehat X_T^A+ \widehat X_T^B)$ is the infimum on
the pairs $(F_T, x) \in\widehat{ \mc  X}$.
\end{lemma}
\textit{Proof.} The result follows observing that any $F_T\in \mc X$
can be rewritten as $F_T= G_T-x1_T$, for some $G_T\in \mc X$ and
$x\in \mathbb{R}$, and
\begin{eqnarray*}
&& {\mathcal R}_{A}\square {\mathcal R}_{B}(X_T^A+ X_T^B) =\inf_{
F_T \in \mc X }
 \big\{
 {\mc R}_A(X_T^A+ X_T^B - F_T)
+ {\mc R}_{B}(F_T) \big\} \\ &&= \inf_{ (G_T,x) \in \mc X \times
\mathbb{R} }
 \big\{
 {\mc R}_A(X_T^A+ X_T^B - (G_T-x1_T))
+ {\mc R}_{B}(G_T-x1_T) \big\}
\\
&& =\inf_{\widehat G_T=(G_T, x) \in \hat {\mc
X}}\big\{\hat{\rho}_A((X_T^A+ X_T^B)1_{\theta=1}- \widehat{G}_T) +
\hat{\rho}_{B}(\widehat{G}_T) \big\} =\hat{\rho}_{A}\square
\hat{\rho}_{B}(\widehat X_T^A+ \widehat X_T^B).\  \   \   \
 \  \  \hfill{\Box}
\end{eqnarray*}
\citeasnoun[Theorem $3.3$]{barrieu_elkaroui_2006}   show that the
inf-convolution of cash additive risk measures is a cash additive
risk measure.
We apply this result to $\hat{\rho}_{A}\square \hat{\rho}_{B}$. When
$\hat{\rho}_{A}\square \hat{\rho}_{B}( 0)
>-\infty$, the inf-convolution  $\widehat{X}\in \widehat{\mc X}
\longmapsto \hat{\rho}_{A}\square \hat{\rho}_{B}(\widehat {X})$  is
a cash additive risk measure\footnote{For the interpretation of the
condition ${\mathcal R}_{A}\square {\mathcal R}_{B}(0)>- \infty$ see
Theorem 3.3 in \citeasnoun{barrieu_elkaroui_2006}.}, continuous from
below if one of the two risk measures is continuous from below, and
with penalty function the sum of the penalties of $\hat{\rho}_{A}$
and $\hat{\rho}_{B}$.
We showed that any  $\hat{\rho}$ constrained to the event $\theta=1$
defines a cash sub-additive risk measure with the same penalty
function (Proposition \ref{pro:hatrhofromR}). Then ${\mathcal
R}_{A}\square {\mathcal R}_{B}$  in equation
(\ref{inf-conv-cashsubadd-cashadd}) is a cash sub-additive risk
measure. We collect all the previous results in the following
theorem.

\begin{theorem}
Let ${\mathcal R} _{A}$ and ${\mathcal R} _{B}$ be two cash
sub-additive risk measures with penalty functions $\alpha _{A}$ and
$\alpha _{B}$, respectively. Let $\mc R _{A,B}$ be the
inf-convolution of $\mc R_{A}$ and $\mc R_{B}$
\begin{eqnarray}
\Psi \rightarrow {\mc R}_{A,B}( \Psi ) :={\mathcal R}_{A}\square
{\mathcal R}_{B}( \Psi ) =\inf_{H\in \mathcal{X}}\big\{{\mathcal
R}_A(\Psi -H) + {\mathcal R}_{B}(H) \big\}
\end{eqnarray}
and assume that ${\mathcal R}_{A,B}( 0) >-\infty$.
Then\\
1) $\mc R _{A,B}$ is a cash sub-additive risk measure which is
finite for all $\Psi \in \mathcal{X}$.\\
2) The associated penalty function is given by \,\,$ \forall
\mathbf{\mu \in }\mathcal{M}_{s,f}(\mc F_T), \,\,\mathcal{\quad
}\alpha _{A,B} (\mathbf{\mu}) =\alpha_{A}( \mathbf{\mu}) +\alpha
_{B}(\mathbf{\mu})$.\\
3) $\mc R _{A,B}$ is continuous from below when this property holds for $\mc R _{A}$ and/or $ \mc R_{B}$.\\
4) The optimal derivative contract is $H^*= F^*- X_T^B$,  where
$F^*$ attains the infimum in (\ref{inf_conv_csarm}).
\end{theorem}

\section{Dynamic infinitesimal cash sub-additive risk measures}
\label{sec:infinitesimal of cash sub-additive risk measures}
The cash sub-additive risk measures considered so far are static
measures assessing the risk of the future position $X_T$ at a given
time $t$.
In this section, we give an example of dynamic cash sub-additive
risk measure on the filtered probability space $(\Omega, \mc
F_T,(\mc F_t)_{t\in [0,T]}, \mathbb{P})$, where $(\mc F_t)_{t\in
[0,T]}$ is the augmented filtration associated to the
$d$-dimensional Brownian motion $W =(W_t)_{t\in[0,T]}$.
At any time $t\in[0,T]$, the risk measure assesses the
riskiness of the future position $X_T$ taking into account the
information available, $\mc F_t$.
In particular, following  \citeasnoun{peng_2004},
\citeasnoun{elkaroui_peng_quenez_1997},
\citeasnoun{barrieu_elkaroui_2006} and
\citeasnoun{rosazzagianin_2006} who link backward stochastic
differential equations (BSDEs) and risk measures, we show that
BSDEs with suitable coefficients are cash sub-additive
risk measures.
The main difference with cash additive risk measures generated by
BSDEs is that cash sub-additive risk measures are now recursive risk
measures, that is the generator can  locally depend on the level of
the cash sub-additive risk measure.
When the dual representation exists, the penalty function of dynamic
cash sub-additive risk measures generalizes the penalty function of
the static cash sub-additive risk measures in
Section~\ref{Examples}.

Dynamic risk measures not based on BSDEs have been  recently studied
by several authors such as \citeasnoun{cvitanic_karatzas_1999},
\citeasnoun{wang_1999},
\citeasnoun{artzner_delbaen_eber_heath_ku_2004}
\citeasnoun{cheridito_delbaen_kupper_2004},
\citeasnoun{frittelli_gianin_2004}, \citeasnoun{riedel_2004},
\citeasnoun{frittelli_scandolo_2005},
\citeasnoun{cheridito_delbaen_kupper_2006}, \citeasnoun{weber_2006}
and \citeasnoun{kloeppel_schweizer_2006}.
Here we consider cash sub-additive risk measures generated by BSDEs.

\subsection{Some results on  BSDEs}
Let $X_T\in L^{\infty}(\Omega, \mc F_T, \mathbb{P})$ and  $g(t, y,
z)$ be a $\mc P\otimes\mc B(\mathbb{R})\otimes\mc
B(\mathbb{R}^d)$-measurable coefficient, where $\mc P$ is the
$\sigma$-algebra of real-valued progressively measurable events.
Consider the pair of squared-integrable progressively measurable
processes $(Y,Z):=(Y_t,Z_t)_{t\in[0,T]}$ solution of the following
BSDE associated to $(g, X_T)$,
\begin{eqnarray*}
-d{Y}_t =  g(t,Y_t, Z_t)dt  - \langle Z_t,dW_ t \rangle, \quad Y_T =
X_T.
\end{eqnarray*}
The existence and the uniqueness of the solution
$(Y_t,Z_t)_{t\in[0,T]}$  depend on the properties of the coefficient
$g$.
\citeasnoun{pardoux_peng_1990} prove that the  solution exists and
is unique  when $g$ is uniformly Lipschitz continuous with respect
to $(y, z)$. In this case $g$ is called standard coefficient.
When, for any given $t\in[0, T]$, $g$ is continuous with respect to
$(y,z)$ $\mathbb{P}$-a.s.\ and  $|g(t,y,z)|\leq C(1+z^2+y),$
$\forall (t,y,z) $ $\mathbb{P}$-a.s., ($g$  with linear-quadratic
growth, in the sequel), \citeasnoun{kobylanski_2000} and
\citeasnoun{lepeltier_sanmartin_1998} show that the  BSDE associated
with $(g, X_T)$ has a maximal and minimal solution. Uniqueness holds
under some additional assumptions.

The following theorem, called \textit{Comparison Theorem}, is
a crucial tool in the study of one-dimensional BSDEs and
corresponding dynamic measures of risk.
\begin{theorem}\label{comparison Theorem}
Let $X_T^1 \mbox{ and } X_T^2\in L^{\infty}(\Omega, \mc F_T,
\mathbb{P})$ and $g^1$ and $g^2$  both standard (or  both with
linear-quadratic growth) coefficients. Let $(Y^1,Z^{1})$ and
$(Y^2,Z^{2})$ be the (maximal) solutions associated  to
$(g^1,X^{1}_T)$ and $(g^{2}, X^{2}_T)$, respectively.
If $ X^{1}_T\geq X^{2}_T$, $\mathbb{P}$-a.s., and
$g^{1}(t,Y^2_t,Z^2_t) \geq g^{2}(t,Y^2_t, Z^2_t) $ $d{\mathbb{P}}
\times dt$-a.s., then $Y^1_t \geq Y^2_t \; \; \mbox{a.s.} \; \forall
\, t\in[0,T]$. In particular, the maximal solution is still monotone
decreasing with respect to the terminal condition.
\end{theorem}
The comparison theorem and the existence of the maximal solution
ensure that, if the coefficient $g$ is convex,  the solution $Y_t$
of the BSDE $(g, -X_T)$ is also convex when $Y_t$ is considered as a
functional of its terminal condition $-X_T$. Moreover, the existence
of the maximal solution ensures the \textit {time consistency} of
${(Y_t)}_{[0, T]}$, that is: $ \forall\,\, 0 \leq t_1 \leq  t_2 \leq
T, \ \ Y_{t_1}(X_T)= Y_{t_1}(-Y_{t_2}(X_T))$. For the derivations of
this result see, for instance,
\citeasnoun{elkaroui_peng_quenez_1997}, \citeasnoun{peng_2004},
\citeasnoun{barrieu_elkaroui_2006} and
\citeasnoun{rosazzagianin_2006}.

\subsection{BSDEs and cash sub-additive risk measures}
The link between measures of risk and BSDEs is particularly
interesting because it enhances interpretation and tractability of
risk measures.
\citeasnoun{barrieu_elkaroui_2006} point out that the coefficient
$g$ of BSDEs can be interpreted as infinitesimal risk measure over a
time interval $[t,t+dt]$ as $\mathbb{E}_{\mathbb{P}}[dY_t |
\mc{F}_t]=-g(t,Y_t,Z_t)dt$ where $Z_t$ is the local volatility of
the conditional risk measure, $\mathbb{V}(dY_t|
\mathcal{F}_t)=|Z_t|^2dt.$
Choosing carefully the coefficient $g$ enables to generate
$g$-conditional risk measures that are locally compatible with
the different agent beliefs.
\begin{example}\label{ex:local_ambiguous_interst_rates}\textsf{Ambiguous interest rates.}
{\rm Assume that locally $\mathbb{E}_{\mathbb{P}}[-dY_t | \mc{F}_t]$
is driven by the worst case scenario generated by an ambiguous
discount rate $\beta = (\beta_t)_{t\in[0,T]}$, where $\beta$ is an
adapted process ranging between two adapted and bounded processes
$(r_t)_{t\in[0,T]}$ and $(R_t)_{t\in[0,T]}$, that is
\begin{equation*}
\mathbb{E}_{\mathbb{P}}[-dY_t^{r,R} | \mc{F}_t]=\sup_{0\leq r_t\leq
\beta_t\leq R_t}(-\beta_t Y_t^{r,R})dt.
\end{equation*}
$Y^{r,R}$ is the first component solution of the BSDE
\begin{equation*}
-dY_t=-\big(r_t Y^+_t- R_t Y_t^-\big)dt-\langle Z_t,dW_t \rangle,
\quad Y_T=-X_T,
\end{equation*}
where $y^+=\sup(y,0)$ and $y^-=\sup(-y,0).$
More precisely, since $(r_t)_{t\in[0,T]}$ and $(R_t)_{t\in[0,T]}$
are assumed to be bounded, $(Y^{r,R}, Z^{r,R})$ is the unique
solution of the standard BSDE with convex Lipschitz coefficient
\begin{equation}\label{eq:coeffgrR}
g^{r,R}(t,y)= R_t y^- - r_t y^+=\sup_{ r_t \leq \beta_t \leq
R_t}(-\beta_t y).
\end{equation}
Notice that $y \mapsto g^{r,R}(t,y)$ is a monotone non increasing
function.
To provide the intuition on this risk measure, we apply the
comparison theorem to the coefficients $g^{r,R}(t,y)$ and
$g(t,y)=(-\beta_t y)$, $\beta_t \in [r_t, R_t]$, with the same
terminal condition $-X_T$.
Since $g^{r,R}(t,y)\geq (-\beta_t y)$,  $Y_t^{r,R}\geq Y_t^{\beta}$
where $Y^{\beta}$ is the solution of the linear BSDE
\begin{equation*}
-dY_t=-\beta_t Y_tdt-\langle Z_t,dW_t \rangle, \quad Y_T=-X_T,
\end{equation*}
and it can be represented as
$Y^{\beta}_t=\mathbb{E}_{\mathbb{P}}[e^{-\int_t^T \beta_s ds}(-X_T)|
\mc{F}_t]$, $\forall t\in[0,T]$.
Then it follows that $Y_t^{r,R}\geq {\rm ess}\sup_{0\leq r_t\leq
\beta_t\leq R_t} Y_t^{\beta}$.
As the process $\overline{\beta}_t=R_t 1_{Y_t^{r,R} \leq
0}+r_t1_{Y_t^{r,R}> 0} $ achieves the maximum of $\sup_{r_t \leq
\beta_t \leq R_t}(-\beta_t Y_t^{r,R})= - \overline{\beta}_t
Y_t^{r,R}$, then the equality $Y_t^{r,R}=Y_t^{\overline{\beta}_t}$
holds. Thus, the dual representation of $Y_t^{r,R}$ follows
\begin{equation*}
Y_t^{r,R}= Y_t^{\overline{\beta}_t} = {{\rm ess}\sup}_{0\leq r_t\leq
\beta_t\leq R_t}\mathbb{E}_{\mathbb{P}}[e^{-\int_t^T \beta_s
ds}(-X_T)| \mc{F}_t].
\end{equation*}
Notice that, for any $t\in[0, T]$,  $Y_t^{r,R}$  is dominated, but
in general not equal to the conditional risk measure $\mc R_t^{
D^R,D^r}$ associated with the worst case discounted factors
$D_{t,T}^R \leq D_{t,T} \leq D_{t,T}^r $, where $D_{t,T}^R=
\exp\{-\int_t^T R_s ds\}$ and $D_{t,T}^r= \exp\{-\int_t^T r_s ds\}$,
\begin{equation}
Y_t^{r,R}(-X_T)\leq \mc R_t^{D^R,D^r}(X_T)
= \mathbb{E}_{\mathbb{P}}[D_{t,T}^R(-X_T)^- +D_{t,T}^r (-X_T)^+|
\mc{F}_t].
\end{equation}
$\mc R^{ D^R,D^r}:=(\mc R_t^{ D^R,D^r})_{t\in[0,T]}$ is a cash
sub-additive risk measure which is not time consistent in contrast
to $Y^{r,R}=(Y^{r,R}_t)_{t\in[0, T]}$.}
\end{example}

In the sequel we consider risk measures generated by BSDEs which
generalize Example \ref{ex:local_ambiguous_interst_rates}.
For the remain part of the paper  $g(t,y,z)$ denotes a convex
generator in $(y, z)$, standard or with linear growth with respect
to $y$ and quadratic growth in $z$. The comparison theorem ensures
that the (maximal) solution $(Y, Z)$ associated with a $(g, -X_T)$
exists and, for any $t\in[0, t]$, $Y_t$ is convex and  decreasing
with respect to the final condition $-X_T$.

The coefficient $g^{r,R}(t,y)$ in equation (\ref{eq:coeffgrR})
depends on $y$ in a convex decreasing way.
As observed by \citeasnoun{peng_2004} and
\citeasnoun{barrieu_elkaroui_2006}, this is never the case for
conditional cash additive risk measures generated by BSDEs.
Under some mild additional assumptions, {\citeasnoun{peng_2004}
shows that, for any $t\in[0, T]$, the (maximal) solution $Y_t$
associated with $(g, -X_T)$ is cash additive as functional of its
terminal condition if and only if $g$ does not depend on $y$ for any
$t\in [0, T]$.
\citeasnoun{barrieu_elkaroui_2006} study these cash additive
solutions  as a dynamic risk measure $(\rho_t(X_T))_{t \in[0, T]}$,
$\rho_t(X_T)= Y_t(-X_T)$, that they  call $g$-conditional risk
measures\footnote{If $g(t, 0)=0$ for any $t\in [0, T]$, the
$g$-conditional risk measures coincide with the non linear
expectation originally studied by \citeasnoun{peng_2004}; see also
\citeasnoun{rosazzagianin_2006}.}.

In the following proposition we show that conditional risk measures
generated by BSDEs are cash sub-additive when  the convex
coefficient $g(t,y,z)$ depends on both $y$ and $z$ and  is
decreasing with respect to $y$.
%
\begin{proposition}\label{pro:dynamicBSSES}
If the convex $g(t,y,z)$ is decreasing with respect to $y$ then
the (maximal) solution $Y_t$ of the BSDE associated with $(g,-X_T)$
is a conditional cash sub-additive risk measure, $\mc R^g_t(X_T)=
Y_t$ and $\mc R^g= (\mc R^g_t(X_T))_{t\in [0, T]}$ is a time
consistent cash sub-additive risk measure. We call $\mc R^g= (\mc
R^g_t(X_T))_{t\in [0, T]}$ \textup{$g$-conditional cash sub-additive
risk measure}.
\end{proposition}
\textit{Proof.} For the convexity and the decreasing monotonicity of
$Y_t$ with respect to the terminal condition see, for instance,
 \citeasnoun{elkaroui_quenez_1996} and  \citeasnoun{peng_2007}. \\
\textit{Cash sub-additivity:} Consider the BSDE satisfied by $\mc
R^g_t(X_T+m1_T)+m= Y_t^m$,
\[
-dY^m_t = g^m \big( t,Y_{t}^m, Z^m_t\big) dt -\langle
Z^m_{t},dW_{t}\rangle,\qquad Y_{T}^m =- X_T.
\]
Since  $g^m(t,y,z)= g(t,y-m,z)$, then $g^m(t,y,z)$ is increasing in
$m$ (as $g$ is decreasing in $y$). From the comparison theorem it
follows that $\mc R^g_t(X_T+m1_T)+m= Y_t^m$ is increasing in $m$.
\hfill{$\Box$}

\subsection{Dual Representation}
In this section we focus on a dual representation for
$g$-conditional cash sub-additive risk measures $\mc R^g$ as in the
static case.
For the cash additive  $g$-conditional risk measures such a
representation has been derived in
\citeasnoun{barrieu_elkaroui_2006}.
The next result is a straightforward generalization of their
results.

The key tool to obtain dual representations is the Legendre
transform of the generator $g$ defined by
\begin{equation*}
G(t, \beta, \mu):=\sup_{(y,z)\in\mathbb{ R} \times
\mathbb{R}^d}\{-\beta y - \langle \mu, z\rangle-g(t, y, z)\}.
\end{equation*}
The following lemma summarizes the properties of $G$ and $g$.
\begin{lemma}\label{lemma:Fenchelduality}
Let $g$ be a continuous convex function on
$\mathbb{R}\times\mathbb{R}^d$ satisfying  the growth control: there
exist two positive constants $C>0$ and $k>0$ such that
$|g(t,y,z)|\leq |g(t,0,0)|+C |y|+\frac k2|z|^2.$
\begin{itemize}
\item[i)] Then the Legendre transform of $g$, $G(t,\beta, \mu)$, takes
infinite values if $\beta \notin [0, C]$. Moreover,
\begin{equation}\label{inequality_i}
G(t, \beta, \mu)\geq  -|g(t,0,0)|+\frac 1{2k}|\mu|^2.
\end{equation}
\item[ii)] Since $g$ is continuous, for any $t\in [0, T]$,
$g(t, Y_t, Z_t)=\sup_{\beta,\mu}\{-\beta_t Y_t- \langle \mu_t,
Z_t\rangle-G(t, \beta_t, \mu_t)\}$. The maximum is achieved at
$(\overline{\beta}_t, \overline{\mu}_t)$ with
$0\leq\overline{\beta}_t\leq C$ and $|\overline{\mu}_t|^2\leq
A\big(|g(t,0,0)|+C|Y_t|\big)+B|Z_t|^2$, for some $A$ and $B$
positive constants.
\end{itemize}
\end{lemma}
\textit{Proof.} \textit{i)} $G(t, \beta, \mu)\geq -\beta y-g(t, y,
0)\geq -\beta y -|g(t,0,0)|-C |y|$. Then, if $|\beta|>C$,
$\sup_{y\in \mathbb{R} }\{-\beta y -C |y|\}= +\infty$.
Moreover, since $g(t, y, z)$ is monotone decreasing with respect to
$y$, $-g(t, y, 0)\geq -g(t, 0,0),$ $\forall y >0$ and $G(t, \beta,
\mu)\geq -\beta y-g(t, 0, 0),$ $\forall y >0$. Then $G(t, \beta,
\mu)=+\infty$ if $\beta<0$.
To prove the inequality (\ref{inequality_i}), we observe that
 $G(t, \beta, \mu) \geq
 \langle \mu, -z\rangle -g(t, 0, z)
 \geq  \langle \mu, -z\rangle -|g(t,0,0)|
- \frac k2 |z|^2$.
As $\max_{z\in \mathbb{R}}\{\langle \mu, -z \rangle - \frac k2
|z|^2\}= \frac1{2k}|\mu|^2$  the result follows.\\
\textit{ ii)} Standard results in convex analysis show that, since
$g$ is continuous, the duality between $g$ and $G$ holds true and
the maximum is achieved.

To show the inequality in \textit{ii)}, we choose a constant $\ve$
such that $0<\ve <\frac 1{2k}$ and we use the inequality
(\ref{inequality_i}),
\begin{eqnarray*}
\big(\frac 1{2k}-\ve\big)|\overline{\mu}_t|^2 &\leq& |g(t,0,0)|+
G(t, \overline{\beta}_t, \overline{\mu}_t)-\ve|\overline{\mu}_t|^2\\
&\leq& |g(t,0,0)|-\overline{\beta}_t Y_t+ \langle\overline{\mu}_t,
-Z_t\rangle-g(t,Y_t,Z_t) -\ve|\overline{\mu}_t|^2\\
&\leq& 2|g(t,0,0)|+2 C|Y_t|+ \frac k{2} |Z_t|^2+
\sup_{\mu_t}\{\langle \mu_t, -Z_t\rangle - \ve |\mu_t|^2\}.
\end{eqnarray*}
As $\max_{\mu_t \in \mathbb{R}}\{\langle \mu_t, -Z_t\rangle - \ve
|\mu_t|^2\}= \frac{|Z_t|^2}{4\ve}$, then
$\big(\frac 1{2k}-\ve\big)|\overline{\mu}_t|^2 \leq2|g(t,0,0)|+2
C|Y_t|+\big(\frac k 2+\frac1{4\ve}\big)|Z_t|^2$, which proves the
inequality. \hfill{$\Box $}

Now we introduce the class of probability measures that appears in
the dual representation. As in \citeasnoun{barrieu_elkaroui_2006}
the reference is the Girsanov theorem for the BMO-exponential
martingales such as defined in \citeasnoun{kazamaki_1994},
\begin{equation*}
\Gamma_t^{\mu}=\mc E(M_t^{\mu})=\exp\big(-\int_0^t \mu_s
dW_s-\frac12 \int_0^t|\mu_s|^2 ds\big),
\end{equation*}
where $M_t^{\mu}=\int_0^t \mu_s dW_s$ is a
BMO($\mathbb{P}$)-martingale, that is $\mu$ belongs to
BMO($\mathbb{P}$),
\[\mbox{BMO}(\mathbb{P}):=\{\psi \in \mc H^2 \mbox{ such that } \exists C>0:
\mathbb{E}_{\mathbb{P}}[\int_t^T \psi_s^2 ds |\mc F_t]\leq C\ a.s.,
\forall t\in[0, T]\},
\]
where $\mc H^2=\{\psi  \in \mc P_1 \mbox{ such that }
\mathbb{E}[\int_0^T ¦\psi ¦^2ds]<\infty \}$.
Using \citeasnoun[Section 3.3]{kazamaki_1994}, $\Gamma_T^{\mu}$ is
the likelihood of an equivalent probability measure on $\mc F_T$
with respect to $\mathbb{P}$ defined by $d\mathbb{Q}^{\mu}=
\Gamma_T^{\mu} d\mathbb{P}$.
Moreover, if $v\in$ BMO($\mathbb{P}$) then $v\in$
BMO($\mathbb{Q}^{\mu}$).
Recall that $\Gamma_t^{\mu}$ is the solution of the forward
stochastic differential equation
\begin{equation*}
d\Gamma_t^{\mu}=\Gamma_t^{\mu}\langle -\mu_t, dW_t\rangle, \quad
\Gamma_0^{\mu}=1.
\end{equation*}
Now we establish  the duality theorem.
\begin{theorem}
Let g be a convex coefficient, decreasing with respect to $y$ and
with growth $|g(t,y,z)|\leq |g(t,0,0)|+C |y|+\frac k 2|z|^2.$
Moreover, assume that there exists a constant $K>0$ such that
$\mathbb{E}\big[ \int_t^T |g(s,0, 0)|ds|\mc F_t\big]\leq K$,
$\forall t\in[0, T]$. Then the (maximal) solution $(Y, Z)$ of the
BSDE
\begin{equation*}
-dY_t = g(t,Y_t,Z_t)-\langle Z_t,dW_t \rangle, \quad Y_T=-X_T, \quad
X_T\in L^{\infty}(\mathbb{P}),
\end{equation*}
is bounded and  $Z$ is in BMO($\mathbb{P}$).
Let $G(t,y,z)$ be the Fenchel transform of $g$ and
\begin{equation*}
\mc A:= \left\{ (\beta_t, \mu_t)_{t\in[0, T]}|\, G(t,\beta_t,\mu_t)<
+\infty,\, 0 \leq \beta_t \leq C,\, \forall t \in [0, T] \mbox{ and
} \mu \in \mbox{ BMO}(\mathbb{P})\right\}.
\end{equation*}
Then, the $g$-conditional cash sub-additive risk measure $\mc R^g=
(\mc R^g_t(X_T))_{t\in [0, T]}$, $\mc R^g_t(X_T)= Y_t$, has the
following dual representation
\begin{eqnarray}
\quad \mc R_t^g(X_T) = {{\rm ess}\sup}_{(\beta,\, \mu) \in \mc A}
{\mathbb{E}}_{\mathbb{Q}^{\mu}} \Big[ e^{-\int_t^T \beta_s ds} \,
(-X_T) - \int_t^T e^{ - \int_t^s \beta_u du} G(s, \beta_s, \mu_s)ds
\big|{\mc F}_t \Big].\label{dual_dynamic_csarm}
\end{eqnarray}
\end{theorem}
\begin{remark}{\rm
The dual representation of $\mc R^g$ in equation
(\ref{dual_dynamic_csarm}) is similar to the dual representation of
static cash sub-additive risk measures. Here, the sub-probability
measures are replaced by the $\mc F_t$-conditional sub-probability
measures $R^{\beta, \mu}$
\begin{equation*}
{\frac{dR^{\beta, \mu}}{d \mathbb{P}}}{|\mc
F_t}:=\exp{\big(-\int_t^T \mu_s dW_s-\frac12 \int_t^T|\mu_s|^2 ds-
\int_t^T \beta_s ds\big)}
\end{equation*}
and the penalty function becomes
\begin{equation*}
\alpha_t(R^{\beta, \mu}):=R^{\beta, \mu}\left(\int_t^T e^{ -
\int_t^s \beta_u du} G(s, \beta_s, \mu_s)ds \big|{\mc F}_t \right).
\end{equation*}}
\end{remark}
\textit{Proof.} \textit{i)} To show that $Z\in \mbox{
BMO}(\mathbb{P}$) we refer the reader to the proof in
\citeasnoun{barrieu_elkaroui_2006}.\\
\textit{ii)} From the Girsanov theorem for BMO-martingales we known
that for any $0 \leq \beta_t \leq C $,  $\mu \in BMO(\mathbb{P})$,
$dW^{\mu}_t= dW_t+ \mu_tdt$ is a $\mathbb{Q}^{\mu}$-Brownian motion
 and
\begin{eqnarray*}
-dY_t&=&g(t,Y_t,Z_t)-\langle Z_t,dW_t \rangle\\&=&
\big[g(t,Y_t,Z_t)+\beta_t Y_t+ \langle \mu_t, Z_t \rangle\big]dt
-\beta_t Y_t dt- \langle Z_t,dW^{\mu}_t\rangle.
\end{eqnarray*}
Then it follows
\begin{eqnarray}\label{ciao_ciao}\nonumber
Y_t(-X_T)&=&\mathbb{E}_{\mathbb{Q}^{\mu}}\Big[ e^{-\int_t^T \beta_s
ds} \, (-X_T) + \int_t^T e^{- \int_t^s \beta_u du}\big[
g(s,Y_s,Z_s)+\beta_sY_s+ \langle \mu_s ,Z_s\rangle \big]ds |\mc
F_t\Big]
\\
&\geq & \mathbb{E}_{\mathbb{Q}^{\mu}}\Big[ e^{-\int_t^T \beta_s ds}
\,(- X_T) - \int_t^T e^{ - \int_t^s \beta_u du} G(s, \beta_s,
\mu_s)ds |\mc F_t\Big].
\end{eqnarray}
To prove the last equality in (\ref{ciao_ciao}) at the optimal
control $(\overline{\beta}, \overline{\mu})$,
\begin{equation*}
G(t,\overline{\beta}, \overline{\mu})= -\overline{\beta}_tY_t-
\langle \overline{\mu}_t, Z_t\rangle-  g(t,Y_t,Z_t), \ \forall
t\in[0, T],
\end{equation*}
we need to verify that $(\overline{\beta}, \overline{\mu})$ is
admissible.
Since $0\leq \overline{\beta}_t\leq C$, we only need  to verify that
$\overline{\mu}$ is in BMO($\mathbb{P})$.
We use the inequality  in Lemma \ref{lemma:Fenchelduality},
$|\overline{\mu}_t|^2\leq A\big(|g(t,0,0)|+c|Y_t|\big)+B|Z_t|^2$.
Since $|g(t,0,0)|^{1/2}$ belongs to BMO($\mathbb{P})$, $Y$ is
bounded and $Z\in$ BMO($\mathbb{P})$, then $\overline{\mu} \in$
BMO($\mathbb{P})$,
\begin{equation*}
Y_t(-X_T)=\mc R^g(X_T)
=\mathbb{E}_{\mathbb{Q}^{\overline{\mu}}}\Big[e^{-\int_t^T
\overline{\beta}_s ds} \,(- X_T) - \int_t^T e^{ - \int_t^s
\overline{\beta}_u du} G(s, \overline{\beta}_s, \overline{\mu}_s)ds
| \mc F_t\Big]
\end{equation*}
and this establishes  the dual representation. \hfill{$\Box$}

\section{Conclusion}\label{sec:conclusion}
We propose a new class of risk measures called cash sub-additive
risk measures which accounts for the risk/ambiguity on interest
rates when assessing the risk of future financial, nonfinancial and
insurance positions.
This goal is achieved by relaxing the debated cash additive axiom
into the cash sub-additive axiom.
We provide several examples of the new risk measures in the static
and the dynamic frameworks, such as the put options premium and the
robust expected utilities. In the dynamic framework cash
sub-additive risk measures are generated by BSDEs enhancing their
tractability and interpretability. 
Cash sub-additive risk measures represent a promising research area
as these risk measures overcome the issues arising from the cash additive axiom.

\newpage
\bibliography{mybiblio}
\end{document}